\documentclass[twocolumn,showpacs,aps,pra,groupedaddress,superscriptaddress]{revtex4-2}

\usepackage{mathrsfs}
\usepackage{amssymb, amsbsy, amsmath, latexsym, dsfont, array, layout, mathrsfs, color}

\usepackage{mathtools}

\usepackage{amsmath,amssymb}
\usepackage{bm}
\usepackage{graphicx}
\usepackage{braket,dsfont}
\usepackage{color}
\usepackage[10pt]{moresize}
\usepackage{mathrsfs}
\usepackage{multirow}
\usepackage{hyperref}
\usepackage{changes}
\usepackage{mathdots}
\let\oldcaption\caption
\renewcommand{\caption}{\sffamily \oldcaption}

\newcommand{\ketbra}[2]{\ket{#1}\!\!\bra{#2}}

\usepackage{amsmath,amsfonts,amssymb}
\usepackage{wrapfig}
\usepackage{graphicx}
\usepackage{bbm}
\usepackage[normalem]{ulem}
\usepackage{color}

\newcommand{\blue}[1]{{\color{blue} #1}}

\usepackage{appendix}
\begin{document}
\title{
Efficient Device-Independent Quantum Key Distribution
}

\author{Shih-Hsuan Chen}
\affiliation{Department of Engineering Science, National Cheng Kung University, Tainan 70101, Taiwan}
\affiliation{Center for Quantum Frontiers of Research and Technology, National Cheng Kung University, Tainan 70101, Taiwan}
\author{Chun-Hao Chang}
\affiliation{Department of Engineering Science, National Cheng Kung University, Tainan 70101, Taiwan}
\affiliation{Center for Quantum Frontiers of Research and Technology, National Cheng Kung University, Tainan 70101, Taiwan}
\author{Chih-Sung Chuu}
\affiliation{Department of Physics, National Tsing Hua University, Hsinchu 30013, Taiwan}
\affiliation{Center for Quantum Science and Technology, Hsinchu 30013, Taiwan}
\author{Che-Ming Li}
\email{cmli@mail.ncku.edu.tw}
\affiliation{Department of Engineering Science, National Cheng Kung University, Tainan 70101, Taiwan}
\affiliation{Center for Quantum Frontiers of Research and Technology, National Cheng Kung University, Tainan 70101, Taiwan}
\affiliation{Center for Quantum Science and Technology, Hsinchu 30013, Taiwan}

\begin{abstract}
Device-independent quantum key distribution (DIQKD) is a key distribution scheme whose security is based on the laws of quantum physics but does not require any assumptions about the devices used in the protocol. The security of the existing entanglement-based DIQKD protocol relies on the Bell test. Here, we propose an efficient device-independent quantum key distribution (EDIQKD) protocol in which one participant prepares states and transmits them to another participant through a quantum channel to measure. In this prepare-and-measure protocol, the transmission process between participants is characterized according to the process tomography for security, ruling out any mimicry using the classical initial, transmission, and final state. Comparing the minimal number of bits of the raw key to guarantee security against collective attacks, the efficiency of the EDIQKD protocol is two orders of magnitude more than that of the DIQKD protocol for the reliable key of which quantum bit error rate is allowed up to $6.5\%$. This advantage will enable participants to substantially conserve the entangled pair's demanded resources and the measurement. According to the highest detection efficiency in the recent most advanced photonic experiment, our protocol can be realized with a non-zero key rate and remains more efficient than usual DIQKD. Our protocol and its security analysis may offer helpful insight into identifying the typical prepare-and-measure quantum information tasks with the device-independent scenario.
\end{abstract}

\maketitle

\section{INTRODUCTION}
\begin{figure}[t]
\includegraphics[width=5.5cm]{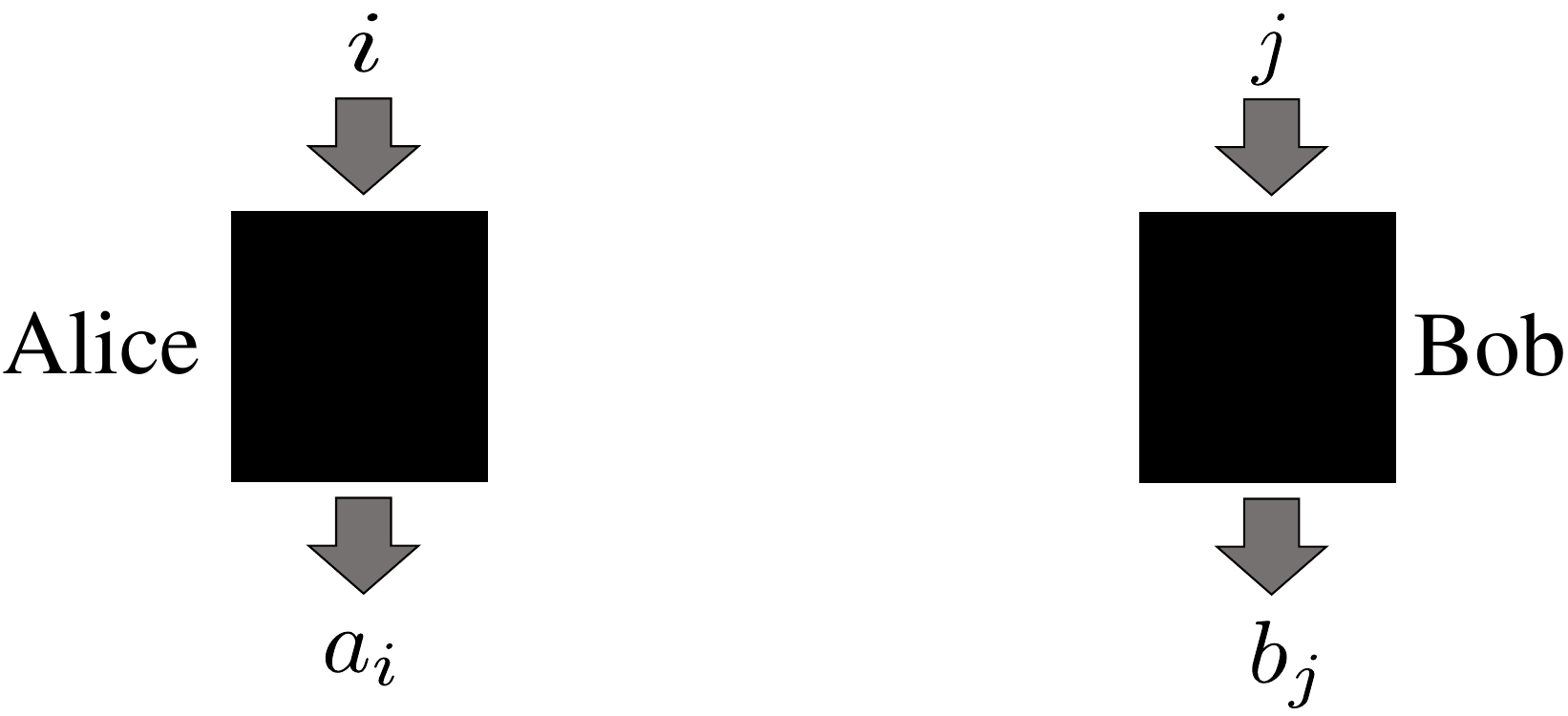}
\caption{The illustration of the DIQKD scenario. In the DIQKD scenario \cite{Pironio09}, Alice's and Bob's devices can be viewed as black boxes. They choose the classical inputs $i$ and $j$ without knowing how devices work and obtain the classical outputs $a_i$ and $b_j$. They can only identify the security of a secret key by the observed classical input-output relations.}
\label{concept_DI}
\end{figure}

Quantum key distribution (QKD) is an approach to generate a cryptographic key between its participants. Unlike classical cryptography of which the security is based on the assumptions of the eavesdropper's computational power, in QKD tasks, the security of the key is based on the laws of physics \cite{gisin2007quantum,lo2014secure,diamanti2016practical,pirandola2020advances}. All QKD schemes rely for security on several assumptions, e.g., a detailed characterization of the devices used to generate the secret key. According to the resources used between sender and receiver, QKD protocols can be classified into two types: entanglement-based QKD protocol \cite{Ekert91} and prepare-and-measure QKD protocol \cite{Bennett84}. The device-independent quantum key distribution (DIQKD) is one entanglement-based QKD protocol \cite{Ekert91} whose security requires minimal assumptions about the devices used in the protocol. In the DIQKD protocol, the secret key is generated by local measurements on an entangled state shared between the participants. The security of the DIQKD protocol is based on observing a Bell-inequality violation, which means that the participants can certify the security of the key with minimal assumptions about the devices used in the protocol \cite{Pironio09,Xu20,Acin07}, i.e., without a detailed characterization of the devices, as shown in Fig.~\ref{concept_DI}.

Another QKD implementation is the prepare-and-measure QKD protocol, e.g., Bennett-Brassard 1984 (BB84) protocol \cite{Bennett84}. In the prepare-and-measure QKD protocol, one participant prepares states and transmits them through a quantum channel to another participant to measure. However, the security of BB84 QKD protocol relies on the assumption that the users' devices are trusted. To our knowledge, there is no security analysis of prepare-and-measure QKD protocol under the device-independent scenario that state-preparation and measurement devices are untrusted or uncharacterized \cite{Xu20, Primaatmaja2023securityofdevice, Zapatero2023}. Therefore, the user of the prepare-and-measure QKD protocol should check the functionality of their devices every time before the QKD tasks and cannot check the devices during the protocol like what DIQKD does.

%Motivated by this present development of prepare-and-measure QKD,
Motivated by this issue, we consider the following questions: Can prepare-and-measure QKD protocol realize secure key distribution under the device-independent scenario? If this is the case, what are the performance differences between it and the usual entanglement-based DIQKD? Herein, we introduce a new DIQKD protocol suitable for being implemented in a prepare-and-measure manner. Unlike Bell-inequality violation, which is used in the DIQKD protocol to distinguish the quantum states from classical states, we use the genuinely classical process (GCP) model~\cite{Chen20} in our protocol to determine quantum processes from classical processes in the whole key generation procedure. A GCP can be described by a process matrix, denoted as $\chi_{\text{GC}}$, that specifies the classical initial state, its evolution, and the final state. This promotes the high efficiency of our prepare-and-measure protocol, which can serve as an efficient DIQKD (EDIQKD) protocol, much better than the usual one.

In the following, we will first introduce the EDIQKD protocol. Then, we provide proof of secure key distribution to show that our method realizes an EDIQKD and compare our protocol with entanglement-based DIQKD through the secure key rates against collective attacks, the required detection efficiency for experimental feasibility, and the required minimum number of key rounds to have a positive key rate. In each round of our protocol, a sender uses the resources of the entangled state, including respective measurements, to prepare the qubit for a receiver. Thus, the less necessary number of rounds, namely the higher efficiency of the protocol, offers the advantage of saving resources. Finally, a conclusion and outlook are given at the end.

\begin{figure}[t]
\includegraphics[width=8.1cm]{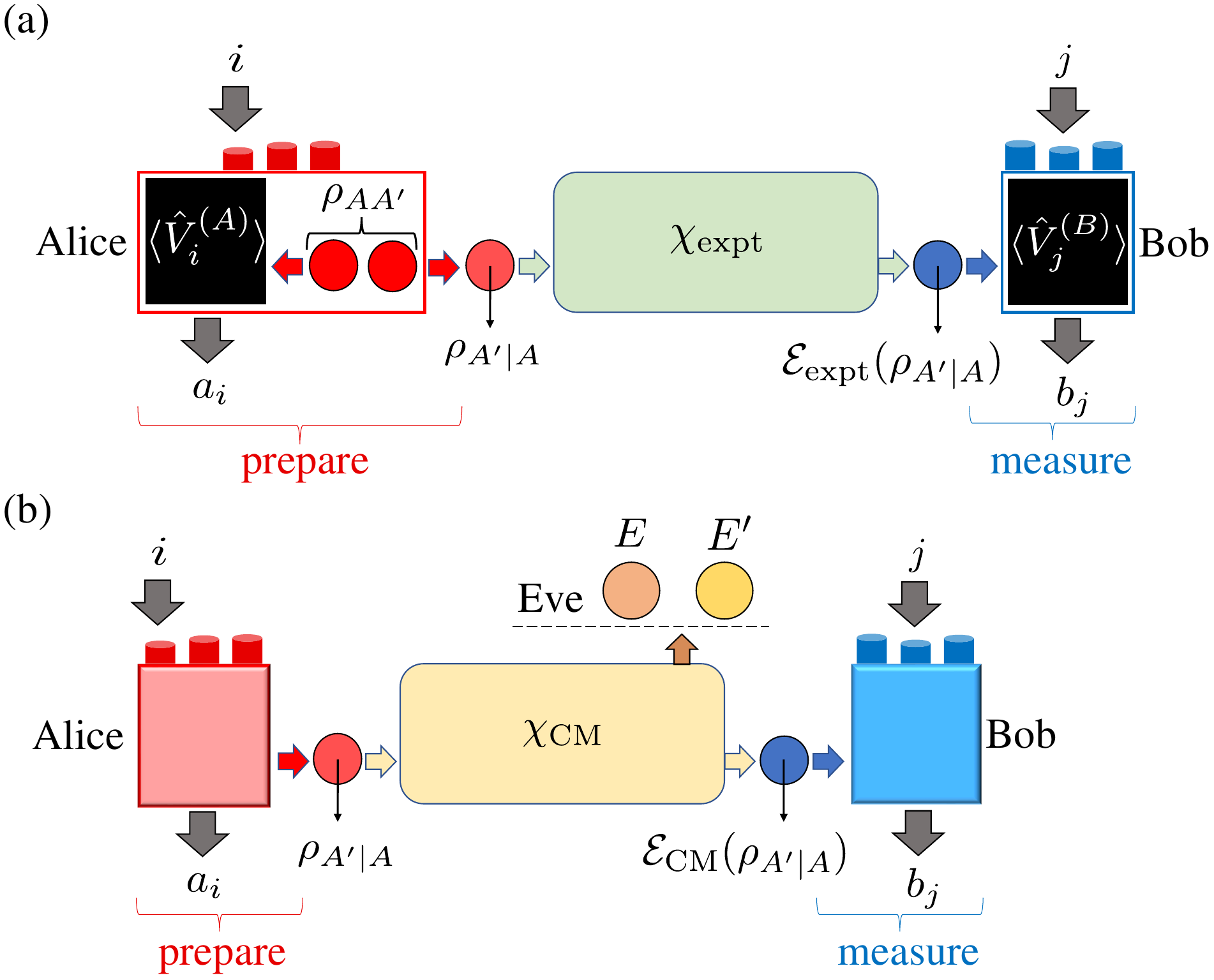}
\caption{The illustration of EDIQKD protocol. (a) In our proposed protocol, we assumed that Alice first prepares the entangled state $\rho_{AA'}$. Then Alice prepares a single quantum state $\rho_{A'|A}$ for Bob by randomly measuring one particle of $\rho_{AA'}$ with an observable $\hat{V}^{(A)}_{i}$, $i \in \{1,2,3\}$, and recording binary outcome $a_i \in \{+1,-1\}$, and she transmits the other particle to Bob. The transmission can be viewed as a single-qubit process $\mathcal{E}_{\rm{expt}}$ which should remain the state in the ideal case, i.e., $\mathcal{E}_{\rm{expt}}(\rho_{A'|A}) = \rho_{A'|A}$. Bob randomly selects an observable $\hat{V}^{(B)}_{j}$, $j \in \{1,2,3\}$, to measure the received state $\mathcal{E}_{\rm{expt}}(\rho_{A'|A})$ and acquire binary outcome $b_j \in \{+1,-1\}$. (b) Since Alice prepares states for Bob to measure, our proposed protocol can be considered the prepare-and-measure scenario. We assume that in the collective attack, Eve uses a cloning machine to get the information of the transmitted qubit $\rho_{A'|A}$ sent from Alice. A tripartite composite system is then created by the cloning machine,
of which the process denoted as $\mathcal{E}_{\rm{CM}}$ is characterized by the process matrix $\chi_{\rm{CM}}$. The state $\mathcal{E}_{\rm{CM}}(\rho_{A'|A})$ is sent to Bob, and the system $E$, together with the ancilla system $E'$, is sent to Eve.}
\label{concept}
\end{figure}

\section{protocol}\label{section:2}

\begin{table}[t]
\caption{EDIQKD protocol for security against collective attacks. In each $k$th round of the EDIQKD experiment, Alice and Bob can randomly choose and record the initial state and measurement bases, respectively, where their choices should be independent. In this work, the total number of rounds is denoted as $N$, the number of key rounds is denoted as $n$, and the length of the final secure key is denoted as $l$. The relation between $N$ and $n$ is $n = N(1-\gamma)$, where $\gamma = (N-n)/N$ is the fraction of test rounds. The relation between $n$ and $l$ is $l=rn$, where $r$ is the key rate.}
\begin{ruledtabular}
\begin{tabular}{lll}
1:\ \textbf{for} $k=1$ to $N$ \textbf{do}\\
2:\ \ \ Alice prepares one eigenstate of $\hat{V}^{(A)}_{i}$ by \\
\ \ \ \ \ \ using her device with the random input $i\in \{1,2,3\}$, and \\
\ \ \ \ \ \ records the binary outcome $a_i \in \{+1,-1\}$.\\
3:\ \ \ Bob chooses one observable $\hat{V}^{(B)}_{j}$ by \\
\ \ \ \ \ \ using his device with the random input $j\in \{1,2,3\}$, and \\
\ \ \ \ \ \ records the binary outcome $b_j \in \{+1,-1\}$.\\
4:\ \ \ Alice and Bob announce their chosen observables.\\
\ \ \ \ \ \ \textbf{if} $i=j=3$ \textbf{then} the measurement result not used\\
\ \ \ \ \ \ in process tomography is the key bit.\\
5:\ \ \ \textbf{else} They announce the results for process \\
\ \ \ \ \ \ tomography algorithm and check whether $F_{\text{expt}}> F_{\text{GC}}$\\
\ \ \ \ \ \ (they aborts if it doesn't meet the criteria).\\
6:\ \ \ end \textbf{if}\\
7:\ end \textbf{for}
\label{protocol}
\end{tabular}
\end{ruledtabular}
\end{table}

In the following, we introduce the protocol of EDIQKD. First, the participants, Alice and Bob, share a quantum channel as shown in Fig.~\ref{concept}(a). In each round for our proposed protocol (step 1 in Table~\ref{protocol}), we assumed that Alice prepares the entangled pair $\rho_{AA'}$. Alice measures one particle of $\rho_{AA'}$ with the observables $\hat{V}^{(A)}_{i}$, $i \in \{1,2,3\}$, chosen randomly by her to prepare the state for Bob, and records the binary outcome $a_i \in \{+1,-1\}$ corresponding to the eigenvalue of $\hat{V}^{(A)}_{i}$ (step 2 in Table~\ref{protocol}). Here, we consider the Pauli-$X$, Pauli-$Y$, and Pauli-$Z$ matrices as Alice's observables for her measurements, i.e., $\hat{V}^{(A)}_{1}=X, \hat{V}^{(A)}_{2}=Y, \hat{V}^{(A)}_{3}=Z$. Notably, the single quantum state is created from the entangled state by Alice, and the detection of one particle heralds the existence of the other one. Given that, Alice knows the state she uses is $\ket{\Psi^{-}}=(\ket{0}\otimes\ket{1}-\ket{1}\otimes\ket{0})/\sqrt{2}$, where $\ket{0}$ and $\ket{1}$ are the eigenstates corresponding to eigenvalues $+1$ and $-1$ of the Pauli-$Z$ matrix, she can correct the prepared states for Bob or his classical outcomes by using the state with its property of anti-correlation. Then she transmits the state $\rho_{A'|A} \in \{\ketbra{0}{0}, \ketbra{1}{1}, \ketbra{+}{+}, \ketbra{-}{-}, \ketbra{R}{R}, \ketbra{L}{L}\}$ to Bob, where $\ket{+}=(\ket{0}+\ket{1})/\sqrt{2}$ and $\ket{-}=(\ket{0}-\ket{1})/\sqrt{2}$ are the eigenstates corresponding to eigenvalues $+1$ and $-1$ of the Pauli-$X$ matrix, $\ket{R}=(\ket{0}+i\ket{1})/\sqrt{2}$ and $\ket{L}=(\ket{0}-i\ket{1})/\sqrt{2}$ are the eigenstates corresponding to eigenvalues $+1$ and $-1$ of the Pauli-$Y$ matrix, and the transmission can be viewed as a single-qubit process $\mathcal{E}_{\rm{expt}}$. Bob randomly chooses one of the observables $\hat{V}^{(B)}_{j}$, $j \in \{1,2,3\}$, to measure his received state $\rho_{B|A} = \mathcal{E}_{\rm{expt}}(\rho_{A'|A})$ and record the binary outcome $b_j \in \{+1,-1\}$ corresponding to the eigenvalue of $\hat{V}^{(B)}_{j}$ (step 3 in Table~\ref{protocol}). Here, we consider that Bob's observables are chosen as $\hat{V}^{(B)}_{1}=UXU^{\dag}, \hat{V}^{(B)}_{2}=UYU^{\dag}, \hat{V}^{(B)}_{3}=UZU^{\dag}$, where $U=\left|0\right\rangle\!\!\left\langle0\right|+\exp(i\pi/4)\left|1\right\rangle\!\!\left\langle1\right|$~\cite{Chen20}.  

In step 4 of Table~\ref{protocol}, Alice and Bob announce the observables they choose and check if those are $\hat{V}^{(A)}_{i}$ and $\hat{V}^{(B)}_{j}$ of which the measurement result not used in step 5 of Table~\ref{protocol} can be the key bit. In step 5 in Table~\ref{protocol}, by analyzing all the measure results in the test rounds to get the probability $P(b_j | a_i)$ of Bob's measurement outcome conditioned on Alice's result $a_i$, Alice and Bob can obtain the density operator of Bob's received state conditioned on Alice's measurement outcome, $\rho_{B|A} = 1/2(\hat{I}+\sum_{j=1}^{3}\sum_{b_j =\pm 1} b_j P(b_j | a_i)\hat{V}^{(B)}_{j})$, where $\hat{I}$ denotes the identity operator, and they can also obtain an analogue of the process matrix $\chi_\text{expt}$ for the channel between them through process tomography (PT) algorithm \cite{Chuang96,Nielsen&Chuang00}
(see Appendix~\ref{appendix:1} for details). Note that the matrix $\chi_\text{expt}$ is not a process matrix in quantum operation formalism, the matrix $\chi_\text{expt}$ will be a process matrix only when the used $\hat{V}^{(A)}_{i}$ and $\hat{V}^{(B)}_{j}$, $i,j \in \{1,2,3\}$, are exactly the observables considered in PT algorithm which relies on trusting the measurement devices. Thus, in the device-independent scenario, the matrix from untrusted devices is only an analogue of the process matrix. To introduce our concept and method fluently, in this work, we still call the matrix $\chi_\text{expt}$ a process matrix.

The ideal EDIQKD protocol acts as an identity unitary transformation $\mathcal{E}_{\hat{I}}$ described by the process matrix $\chi_{I}$ on Alice's transmitted state $\rho_{A'|A}$, i.e., $\mathcal{E}_{\hat{I}}(\rho_{A'|A}) = \rho_{A'|A}$. The security of EDIQKD relies on ruling out all the simulations of classical physics \cite{Chen20}, i.e., from $\chi_{\text{GC}}$. The fidelity of $\chi_\text{expt}$ and $\chi_{I}$ should satisfy
\begin{equation}
F_{\text{expt}}\equiv \text{tr}(\chi_{\text{expt}}\chi_{I})> F_{\text{GC}}\equiv\max_{\chi_{\text{GC}}} \text{tr}(\chi_{\text{GC}}\chi_{I}),\label{fidelity}
\end{equation}
which means $\chi_\text{expt}$ can surpass any classical mimic $\chi_{\text{GC}}$. The classical bound $F_{\text{GC}}$ can be obtained by performing the mathematical maximization task via semi-definite programming with MATLAB \cite{Lofberg, sdpsolver} (see Appendix~\ref{appendix:1} for details). It is worth noting that, in EDIQKD protocol, the security is certified with the process between Alice and Bob. Thus, the input states and the measurement should be prepared according to the PT algorithm. Therefore, compared with the usual entanglement-based DIQKD protocol, which does not characterize the process between participants, i.e., the channel between Alice and Bob in Fig.~\ref{concept}(a), EDIQKD protocol is designed for the prepare-and-measure scenario.

As considered in the usual DIQKD~\cite{Murta19}, an EDIQKD protocol consists of three phases, and the steps described above belong to the first phase, where $n$-bit raw keys are generated. Alice and Bob perform the error correction protocol in the second phase to correct their $n$-bit raw keys and obtain $l'$-bit keys ($l' \leq n$). If the error correction protocol aborts (with the probability less than and equal to ${\epsilon}_\text{EC} $) they abort the EDIQKD protocol, where ${\epsilon}_\text{EC}$ is error probability of the error correction protocol. In the third phase, they perform the privacy amplification protocol to remove an eavesdropper's knowledge about their $l$-bit final secure keys ($l \leq l'$).

In the previous works, there is no security analysis of prepare-and-measure QKD protocol \cite{Bennett84} under the device-independent scenario \cite{Pironio09,Xu20,Acin07}. Since Alice prepares states and transmits them through a quantum channel to Bob to measure, our proposed protocol can be viewed as the prepare-and-measure scenario, as shown in Fig.~\ref{concept}(b). Furthermore, to identify the security of QKD protocol under the device-independent scenario, in our proposed protocol, we assumed that the entangled state is actually exploited in Alice's state preparation for Bob [Fig.~\ref{concept}(a)], and both of their measurement devices are viewed as the black boxes, namely they choose the classical inputs without any assumption about the internal operation of devices and obtain the classical outputs, as shown in Fig.~\ref{concept_DI}. The security can only be identified by the observed statistics. Therefore, our proposed protocol is the first device-independent QKD protocol in the prepare-and-measure scenario with the security analysis from the perspective of the process.

To identify the faithful key against the collective attack, we assume that the experiment rounds are independent and identically distributed (IID), known as IID assumption \cite{Acin07,Pironio09,Xu20,Murta19}. If the preparation and measurement devices are memoryless and the channel between Alice and Bob is the same for every round in the protocol, the IID assumption is satisfied. Furthermore, the IID assumption is also an assumption in PT algorithm \cite{Chuang96}, which is used to get the matrix $\chi_\text{expt}$ describing the distribution $P(b_j | a_i)$ of measurement results $a_i$ and $b_j$ between Alice and Bob in our QKD protocol.

In the collective attacks satisfying the IID assumption, the eavesdropper can make arbitrary global operations on her quantum side information in each round. In contrast, the coherent attacks needn't meet the IID assumption, and the eavesdropper considered in a coherent attack has quantum memory and performs arbitrary global operations according to the results of previous rounds. Thus, the process done by the eavesdropper is related to previous rounds, i.e., the experiments are not memoryless. To analyze the security against coherent attacks in our protocol, we should decompose the total number of rounds $N$ into $m$ blocks with different lengths and verify whether these blocks satisfy IID assumption \cite{Murta19}. For devices satisfy the IID assumption, all $m$ blocks should have similar results.

\section{Discussion}
\subsection{Key rates for collective attacks}\label{section:3_1}
In the collective attack as shown in Fig.~\ref{concept}(b), the eavesdropper, Eve, uses a cloning machine, of which the process denoted as $\mathcal{E}_{\rm{CM}}$ is characterized by the process matrix $\chi_{\rm{CM}}$, acting on the state $\rho_{A'|A}$ sent from Alice, and sends the state $\rho_{B|AE} = \mathcal{E}_{\rm{CM}}(\rho_{A'|A})$ to Bob. We consider that a cloning machine used as an eavesdropping attack is the universal quantum copying machine (UQCM) \cite{Masab22}, of which the output cloned states are independent (or universal) to the input state with the high state fidelity $5/6 \sim 83.33\%$. (The problem of estimating Eve's information is equivalent to finding her optimal eavesdropping attack, which is consistent with quantum mechanics \cite{Scarani05}. In collective attacks, Eve's optimal attack can be identified with an optimal cloner \cite{PhysRevA.85.052310}, where Eve's copy system $E$ comes out with maximal average state fidelity $5/6$. Besides, a composite system among Alice, Bob, and Eve, created by the cloning machine \cite{Chiu16}, can be discussed for security against collective attacks.)
Then Eve collects the system $E$ and the ancilla system $E'$ to gain information about Alice's and Bob's measurement outcomes. Without loss of generality, UQCM used by Eve is assisted by the ancilla system $E'$, which does not contain any information about Alice's input state \cite{Masab22}. In the following, we derive the key rate, which used criteria Eq.~(\ref{fidelity}) for certifying security under the assumption of collective attacks.

The key rate $r$ is the ratio of the final secure key length $l$ in the key rounds $n$. For collective attacks \cite{Acin07}, the infinite key rate $r$ is lower bounded by the Devetak-Winter rate \cite{Devetak05},
\begin{equation}\label{DKkeyrate}
r\geq I(A:B)-I(A:E),
\end{equation}
which is the difference between the mutual information between Alice and Bob, $I(A:B)=1-h(Q)$ [$h$ is the binary entropy and $Q$ is quantum bit error rate (QBER)], and the mutual information between Alice and Eve, $I(A:E)$.
After operating the cloning machine \cite{Scarani05} on a single system sent from Alice, the state in the composite system of Bob and Eve \cite{Chiu16} is
\begin{equation}\label{QCM}
\ket{\phi}_{BEE'}=\sum_{j,k=0}^{1}\sqrt{\lambda_{jk}}\ket{\phi_{jk}}_B\ket{\phi_{j,2-k}}_{EE'},
\end{equation}
where $\lambda_{jk}$ is the probability of observing $\ket{\phi_{jk}}_B$ and $\ket{\phi_{j,2-k}}_{EE'}$,
$\ket{\phi_{jk}}_B = U_{jk}\ket{s}_{Bi}$,
$U_{jk} = \sum_{s=0}^{1}\text{exp}(i\pi sk)\ket{s\oplus j}_{11}\bra{s}$ and $\oplus$ denotes addition module 2,
$\{\ket{s}_{1} = \ket{s}_{B1}\}$ is an orthonormal basis that corresponds to a basis of the first Bob's measurement,
$\ket{s}_{Bi}$ is an orthonormal basis that corresponds to a basis of the $i$-th Bob's measurement, and
$\ket{\phi_{j,2-k}}_{EE'} = (\hat{I} \otimes U_{jk})(1/\sqrt{2})\sum_{s=0}^{1}\ket{s}_{1}\otimes\ket{s}_{1}
= (1/\sqrt{2})\sum_{s=0}^{1}\text{exp}(i\pi sk)\ket{s}_{1}\ket{s\oplus j}_{1}$.
Through IID assumption, the state of Bob's qubit in each round can be represented as $\rho_{B|AE}=\sum_{j,k=0}^{1}\lambda_{jk}\ket{\phi_{jk}}_{BB}\bra{\phi_{jk}}$. With this reduced state, we can obtain the mutual information $I(A:B)$.

To determine the mutual information of Alice's and Eve's measurement results $I(A:E)$, we first consider the mutual dependence between Alice's input and the results derived from measurements on the subsystem composed of Eve's system $E$ and the ancilla system $E'$, as shown in Fig.~\ref{concept}(b). Since Eve's system comprises system $E$ and the ancilla system $E'$, the mutual information $I(A:E)$ is constrained by the Holevo bound \cite{Nielsen&Chuang00},
\begin{equation}\label{eq4}
I(A:E)=S(\rho_{EE'})-\frac{1}{2}\sum_{a_i=\pm1}S(\rho_{EE'|a_i}),
\end{equation}
where $S(\rho_{EE'})=\sum_{j,k=0}^{1}-\lambda_{jk}\text{log}_2\lambda_{jk}$ is the von-Neumann entropy of the state $\rho_{EE'}=\sum_{j,k=0}^{1}\lambda_{jk}\ket{\phi_{j,2-k}}_{EE'EE'}\bra{\phi_{j,2-k}}$, and
$S(\rho_{EE'|a_i})=h(\lambda_{10}+\lambda_{00})$ \cite{Sheridan10} is the von-Neumann entropy of $\rho_{EE'|a_i}$ which denotes the reduced state conditioned on the result $a_i$ obtained by Alice.
Considering a cloner that $\lambda_{00}=(1-p)$ and $\lambda_{01}=\lambda_{10}=\lambda_{11}=p/3$ since the eavesdropping attacks by using UQCM on three orthogonal bases are the same, the key rate $r=0$ while the QBER $Q = 6.9\%$ (as shown in Appendix~\ref{appendix:2}) and the process fidelity $F_{\text{expt}}=0.8656>F_{\text{GC}}=0.8536$.

For the sake of comparison, we compare our key rate with which under the usual DIQKD protocol \cite{Acin06} against collective attacks. In the usual DIQKD protocol, Alice and Bob first share a quantum channel consisting of a source of entangled particles. Then, they use Clauser-Horne-Shimony-Holt (CHSH) inequality \cite{Clauser69} to examine the entangled states. If the entangled pairs are faithful to violate CHSH inequality, the raw key is extracted from measurement results of a specific basis. Considering collective attack, the critical QBER $Q$ is $7.1\%$ \cite{Acin07} (as shown in Appendix~\ref{appendix:2}, where the number of key rounds $n = 10^{15}$ is considered). The critical QBER $Q$ in EDIQKD protocol is stricter than but close to that in the usual DIQKD protocol, which shows that our protocol is comparable in the asymptotic limit of infinite key lengths (for $n \geq 10^{15}$ \cite{Rotem2018}). In the following, we will discuss the secure key rates of protocols in the case of finite key lengths.

\subsection{\label{COSKR}Comparison of secure key rate in finite key lengths}

The study in Ref.~\cite{Acin07} provides the security bound in the asymptotic limit of infinite key lengths. Still, it is not practical for the user to establish a secure key in the experiment and the implementation of QKD. Therefore, resources consumed in the finite number of rounds cause an essential issue of QKD. To compare EDIQKD with entanglement-based DIQKD \cite{Murta19}, we illustrate the variation of the finite key rates $r$ for the number of key rounds $n$ in these protocols, where $l=rn$ is the length of the secure key in $n$ key rounds.

An EDIQKD protocol is secure if it is correct and secret. The QBER of final keys is denoted as $Q^{(l)}$. Correctness means that participants Alice and Bob share the same final secret key, i.e., $Q^{(l)} = 0$, which is ensured by the error correction protocol. Secrecy means that the eavesdropper, Eve, has no information about the final secret key. For the secrecy of DIQKD protocol~\cite{Murta19}, one needs to estimate how far the state describing Alice's final key and the eavesdropper system is from a state where the eavesdropper is unaware of Alice's final key, which is described by the maximally mixed state. Given the usual DIQKD protocol, we propose that, for the secrecy of EDIQKD protocol, one needs to estimate how far the analogue of process matrix in the protocol under eavesdropper's attacks is from an ideal one isolated from eavesdropper's attacks, where this process matrix is considered for three phases of EDIQKD protocol and denoted as $\chi^{(l)}$. If an $l$-bit key is $\epsilon_\text{corr}-$correct and $\epsilon_\text{sec}-$secret, it is secure.

\textbf{Definition (Correctness).}
An $l$-bit final key is $\epsilon_\text{corr}-$correct if the QBER of final keys $Q^{(l)}$ is smaller than and equal to $\epsilon_\text{corr}$, i.e.,
\begin{eqnarray}
Q^{(l)}&\leq& \epsilon_\text{corr}.
\end{eqnarray}

\textbf{Definition (Secrecy).}
An $l$-bit final key is $\epsilon_\text{sec}-$secret if, for every process matrix $\chi_{\text{CM}}^{(l)}$ it holds that
\begin{eqnarray}
D(\chi_{\text{CM}}^{(l)},\chi_{\text{sep}}^{(l)})&\leq& \epsilon_\text{sec},
\end{eqnarray}
where $D$ represents the trace distance of the process matrix in the protocol under collective attacks to an ideal one, $\chi_{\text{CM}}^{(l)}$ is the process matrix characterizing the process of the protocol under eavesdropper's attacks, and $\chi_{\text{sep}}^{(l)}$ is the process matrix characterizing the process of the protocol where the channel between Alice and Bob is isolated from Eve's attack. Furthermore, we explain the input and output states of the processes in detail. $\chi_{\text{CM}}^{(l)}$ characterizes the 2-qubit process $\mathcal{E}_{\text{CM}}^{(l)}$ which contains the use of cloning machine in the protocol of generating $l$-bit final key, where the output state is $\rho_{BE} = \mathcal{E}_{\text{CM}}^{(l)}(\rho_{A'|A}\otimes\ket{0}_{EE}\!\bra{0})$, which represents $\rho_{B|AE}$ together with the system $E$, and $\ket{0}_{EE}\!\bra{0}$ is Eve's initial state.
Besides, $\chi_{\text{sep}}^{(l)} = \chi_{\text{AB}}^{(l)} \otimes \chi_{\text{AE}}^{(l)}$, where
$\chi_{\text{AB}}^{(l)}$ characterizes the 1-qubit process $\mathcal{E}_{\text{AB}}^{(l)}$ of which the output state is $\rho_{B|A} = \mathcal{E}_{\text{AB}}^{(l)}(\rho_{A'|A}) = \rho_{A'|A}$ in the ideal case, and $\chi_{\text{AE}}^{(l)}$ characterizes the 1-qubit process $\mathcal{E}_{\text{AE}}^{(l)}$ of which the output system is $\mathcal{E}_{\text{AE}}^{(l)}(\ket{0}_{EE}\!\bra{0}) = \ket{0}_{EE}\!\bra{0}$ in the ideal case.
The trace distance $D$ for the secrecy in the first phase of the EDIQKD protocol is estimated in Appendix~\ref{trD} as an example for understanding.

The key rate $r$ can be derived from Devetak-Winter rate in Eq.~(\ref{DKkeyrate}).
The mutual information $I(A:B)$ for key rounds and test rounds are $1- h(Q)$ and $1-h(F_{\text{expt}})$, respectively.
Thus $I(A:B) = (1-\gamma)[1- h(Q)] + \gamma[1-h(F_{\text{expt}})] = 1-(1-\gamma) h(Q)-\gamma h(F_{\text{expt}})$
and key rate $r$ can be rephrased as
\begin{eqnarray}\label{keyratef1}
r&\geq& 1-(1-\gamma) h(Q)-\gamma h(F_{\text{expt}})\nonumber\\
&&-[S(\rho_{EE'})-\frac{1}{2}\sum_{a_i=\pm1}S(\rho_{EE'|a_i})]-\frac{\text{leak}_{\text{EC}}}{n},
\end{eqnarray}
where $\gamma$ is the fraction of test rounds and
\begin{eqnarray}\label{keyratef2}
\text{leak}_{\text{EC}}\!&=
&\!\!\sqrt{n}\!\!\left[\!4\text{log}_2(2\sqrt{2}\!+\!\!1)\!\!\left(\!\!\sqrt{\text{log}_2\frac{2}{\epsilon_\text{s}^2}}\!\!+\!\!\sqrt{\text{log}_2\frac{8}{{\epsilon}_\text{EC}'^2}}\right)\!\!\right]\nonumber \\
&&\!\!\!\!+\text{log}_2\!\left(\!\frac{8}{{\epsilon}_\text{EC}'^2\!\!\!}\!+\!\frac{2}{2-{\epsilon}_\text{EC}'}\!\!\right)\!+\!\text{log}_2\!\left(\!\!\frac{1}{\epsilon_\text{EC}}\!\right)\!\!\!\nonumber \\
&&\!\!\!\!+2\text{log}_2\!\left(\!\frac{1}{2\epsilon_\text{PA}}\!\right)\!\!
\end{eqnarray}
is the leakage due to error correction step (see Appendix~\ref{appendix:4} for details of the derivation of key rate in EDIQKD protocol).
$\epsilon_\text{s}$ is the smoothing parameter, and $\epsilon_\text{PA}$ is the error probability of the privacy amplification protocol.

\begin{figure}[t]
\includegraphics[width=7.9cm]{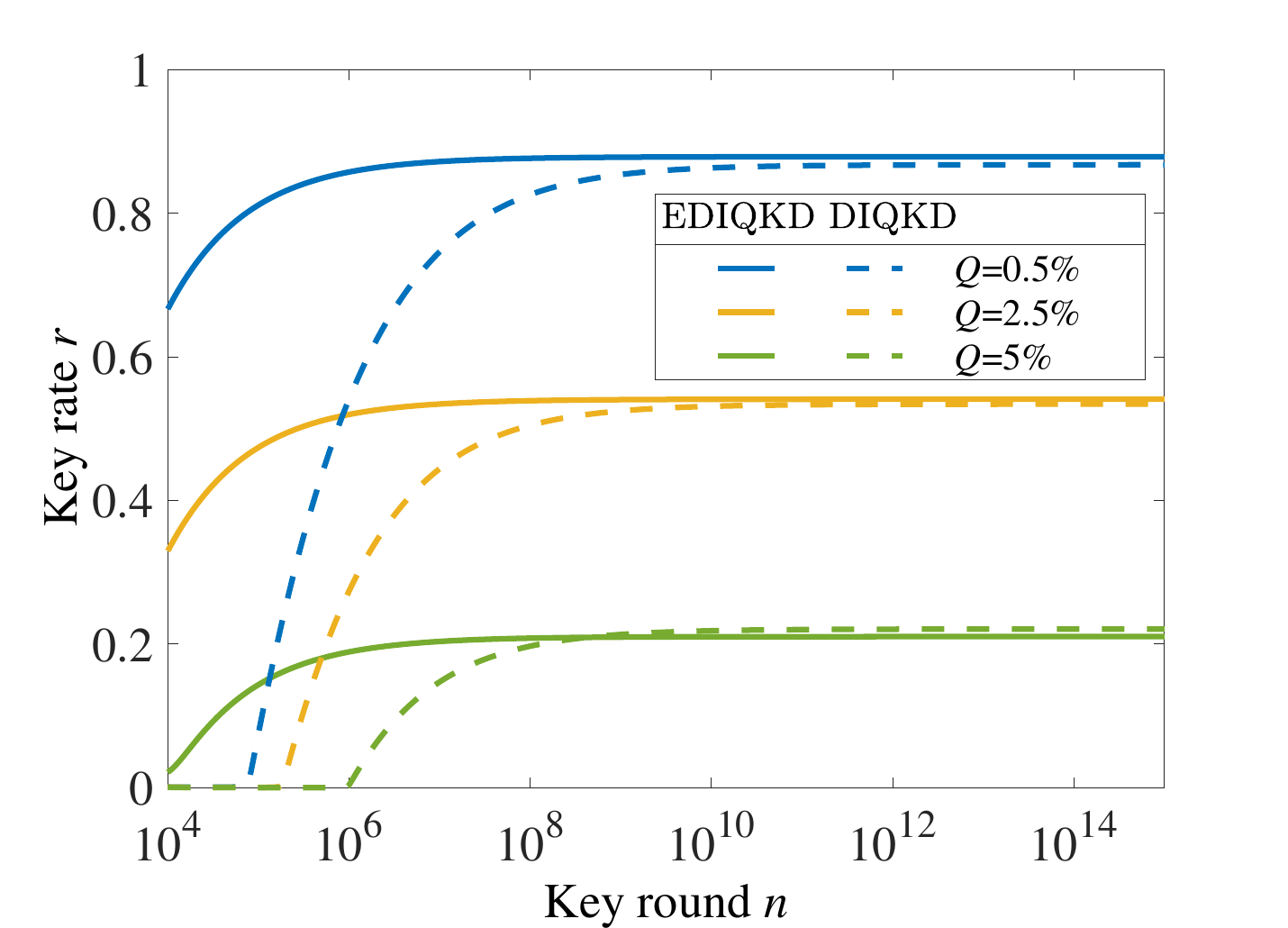}
\caption{Key rate $r$ versus the number of key rounds $n$.
According to the result of Fig.~3 in Ref.~\cite{Murta19}, the QBER $Q$ is chosen as $0.5\%$, $2.5\%$, and $5\%$, and the fraction of test rounds $\gamma$ is set as $10^{-2}$ for the comparison in this study. To compare the key rate in EDIQKD protocol with which in DIQKD protocol~\cite{Murta19}, we consider $\epsilon_\text{s}=10^{-5}$ and $\epsilon_\text{EC}= \epsilon_\text{EC}'=\epsilon_\text{PA}=10^{-2}$ to plot the relation between the key rate $r$ and the number of key rounds $n$ for both protocols. The values, $10^{-5}$ and $10^{-2}$, are chosen as the same values for comparison and the objective standards in the error correction protocol \cite{Murta19}, but there is no broad consensus on their values~\cite{Tomamichel17}.
}\label{keyraten}
\end{figure}

\begin{table}[t]
\caption{Five examples (for the QBER $Q < 6.9\%$) of efficiency factors $E_\text{f}$. $n_{\text{EDIQKD}}'$ ($n_{\text{DIQKD}}'$) is derived from the minimum number of key rounds for key rate $r \geq 0.001$ (chosen as the nonzero key rate in the numerical analysis) in the program data of EDIQKD (DIQKD) protocol. The efficiency factor $E_\text{f}$ is obtained from $n_{\text{DIQKD}}'$ divided by $n_{\text{EDIQKD}}'$.}
\begin{ruledtabular}
\begin{tabular}{llll}
$Q$ & $n_{\text{EDIQKD}}'$ & $n_{\text{DIQKD}}'$ & $E_\text{f} = n_{\text{DIQKD}}' / n_{\text{EDIQKD}}'$\\
[0.2em]
\hline\\
[-0.9em]
$5.5\%$ & $10^{3.77}$ & $10^{6.23}$ & $10^{2.46} \sim 288.40$\\
$6\%$ & $10^{4.18}$ & $10^{6.56}$ & $10^{2.38} \sim 239.88$\\
$6.5\%$ & $10^{5.06}$ & $10^{7.10}$ & $10^{2.04} \sim 109.65$\\
$6.6\%$ & $10^{5.31}$ & $10^{7.26}$ & $10^{1.95} \sim 89.125$\\
$6.7\%$ & $10^{6.14}$ & $10^{7.45}$ & $10^{1.31} \sim 20.42$\\
%$6.8\%$ & $10^{7.51}$ & $10^{7.7}$ & $10^{0.19} \sim 1.55$\\
\end{tabular}
\end{ruledtabular}
\label{fig3table}
\end{table}

Our EDIQKD protocol can be implemented through entangled pairs, which Alice prepares (see Fig.~\ref{concept} for implementation details). As shown in Fig.~\ref{keyraten}, for the same value of QBER $Q$, the key rate $r$ in the EDIQKD protocol is higher than that in the entanglement-based DIQKD protocol. The key rates in both protocols approach similar asymptotic values. Still, the key rates in the EDIQKD protocol rise to the asymptotic values faster than those in the DIQKD protocol within the finite number of key rounds. Furthermore, the minimum number of key rounds required to guarantee security (key rate $r > 0$) in the DIQKD protocol is higher by about two orders of magnitude than that in our protocol; namely, the efficiency of EDIQKD protocol is about 100 times more than that of DIQKD protocol. To objectively compare the efficiency of both protocols, we introduce the efficiency factor $E_\text{f}$ into our work.
The efficiency factor $E_\text{f}$ is defined as the minimum number of key rounds to guarantee security in DIQKD protocol, $n_{\text{DIQKD}}'$, divided by which in EDIQKD protocol, $n_{\text{EDIQKD}}'$, i.e., $E_\text{f} = n_{\text{DIQKD}}' / n_{\text{EDIQKD}}'$. To show how more efficient EDIQKD protocol than DIQKD protocol~\cite{Murta19} is to guarantee security (key rate $r \geq 0.001$ is chosen as the nonzero key rate in the numerical analysis), we take five QBERs $Q < 6.9\%$ as examples and show the corresponding efficiency factors $E_\text{f}$ in Table~\ref{fig3table}. For example, the efficiency of EDIQKD protocol is $10^{2.04} \sim 109.65$ times higher than that of DIQKD protocol for the QBER $Q = 6.5\%$.

The difference between the efficiencies of EDIQKD and DIQKD results from the physical models for security analyses and the measurement settings for algorithms in the protocols.
Here, we compare the combinations of measurement settings in both protocols. There are nine combinations of measurement settings, i.e., $\{(\hat{V}_{i}^{(A)}, \hat{V}_{j}^{(B)})| i, j=1, 2, 3\}$ where $\hat{V}^{(A)}_{1}=X, \hat{V}^{(A)}_{2}=Y, \hat{V}^{(A)}_{3}=Z, \hat{V}^{(B)}_{1}=UXU^{\dag}, \hat{V}^{(B)}_{2}=UYU^{\dag}, \hat{V}^{(B)}_{3}=UZU^{\dag}$, in our protocol, and those for test rounds include that for key rounds. Thus, the measurement settings for key rounds are fully verified.
While there are six combinations of measurement settings in DIQKD protocol \cite{Murta19}, e.g., $\{(\hat{V}_{i}'^{(A)}, \hat{V}_{j}'^{(B)})| i=1, 2\ \text{and } j = 1, 2, 3 \}$,
and $(\hat{V}_{2}'^{(A)}, \hat{V}_{3}'^{(B)})$ is for key rounds.
However, $\{(\hat{V}_{i}'^{(A)}, \hat{V}_{j}'^{(B)})| i, j=1, 2\}$ for test rounds involves $\hat{V}_{2}'^{(A)}$ but doesn't involve $\hat{V}_{3}'^{(B)}$. Furthermore, $(\hat{V}_{1}'^{(A)}, \hat{V}_{3}'^{(B)})$ is chosen for neither key rounds nor test rounds and is an uncontrollable waste of rounds. Therefore, the utilization rate of rounds in the EDIQKD protocol is higher than that in the usual DIQKD protocol, and the EDIQKD protocol is more efficient than the typical DIQKD protocol.

Notably, PT \cite{Chuang96,Nielsen&Chuang00} consumes more measurement settings than the Bell test, but this disadvantage is not revealed in the EDIQKD protocol. In the example above, according to the 1-qubit PT algorithm, the numbers of Bob's measurement settings are the same in both protocols, and Alice's measurement settings in the EDIQKD protocol are just one more than those in the usual DIQKD protocol. Besides, there are wasted measurement settings in the typical DIQKD protocol, but measurement settings are effectively utilized in the EDIQKD protocol. Therefore, compared with the usual DIQKD protocol, the EDIQKD protocol verifies the physical security of keys more directly and effectively.

\begin{figure}[t]
\includegraphics[width=8cm]{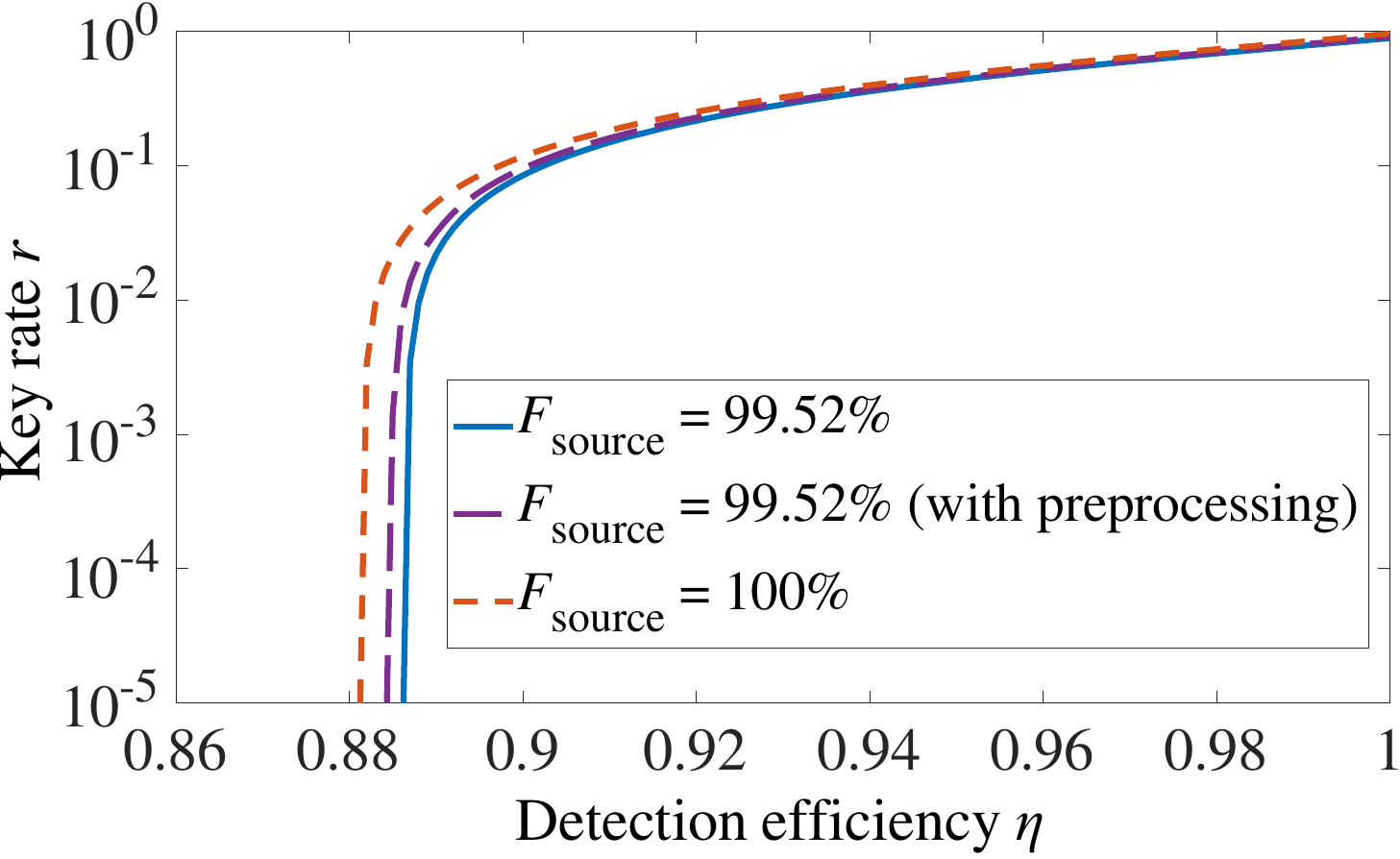}
\caption{
Variation of key rate with different detection efficiencies (key rate $r \geq 10^{-5}$ is chosen as the nonzero key rate in the numerical analysis).
The required detection efficiency for key generation in our protocol is $88.7\%$ with practical imperfections (plotted by blue curve),
and the required detection efficiency drops to $88.1\%$ for an ideal pure state (plotted by orange curve).
With random post-selection and noisy preprocessing~\cite{Liu21}, the required detection efficiency is $88.5\%$ (plotted by purple curve).
The entangled state fidelity $F_{\text{source}}=99.52\%$ is considered here for the practical imperfections~\cite{Liu21}.
}\label{keyratedetecteff}
\end{figure}

\subsection{\label{ESDET}Estimation of required detection efficiency}

Our protocol can be applied and realized to the most advanced photonic experiment in the recent work~\cite{Liu21}, and the security in our protocol can be certified with their experimental platform, which highlights the experimental feasibility of our protocol. Therefore, in Appendix~\ref{appendix:5}, we illustrate the photonic experimental setup in Figs.~\ref{experimentsetup}(a),~\ref{experimentsetup}(b), and~\ref{experimentsetup}(c), which is the same as that in Ref.~\cite{Liu21} and show the required efficiency of devices for our protocol in Fig.~\ref{experimentsetup}(d). In the experiment of Ref.~\cite{Liu21}, the fidelity of the entangled state produced from the spontaneous parametric down-conversion (SPDC) source is $99.52\%$, and the total number of rounds $N = 1.44\times10^9$. To estimate the required detection efficiency for our protocol realized by the photonic experimental setup in Ref.~\cite{Liu21}, we execute the simulation with three critical factors in experimental imperfections \cite{Liu21}, including imperfection of entangled state, the dark count probability $10^{-6}$ of detectors, and the multiphoton effects, and the key rate is optimized under different detection efficiencies by performing the mathematical maximization task with MATLAB.

In the simulations of DIQKD~\cite{Liu21} and EDIQKD protocols, key rates are maximized in the optimization problems, but experimental parameters used to find the optimized key rate are not exactly the same.
In the simulation of DIQKD protocol~\cite{Liu21},
first, it is assumed that the ideally non-maximally entangled state $\ket{\psi}= \cos(\alpha)\ket{0}\otimes\ket{1}+\sin(\alpha)\ket{1}\otimes\ket{0}$ is prepared in each round, where $\alpha \in [0^{\circ}, 45^{\circ}]$.
Second, the observables for Alice's and Bob's measurements are $\hat{V}_{i}'^{(A)} = \cos(\phi_{Ai})Z+\sin(\phi_{Ai})X$ and $\hat{V}_{j}'^{(B)} = \cos(\phi_{Bj})Z+\sin(\phi_{Bj})X$, respectively, where $\phi_{Ai}, \phi_{Bj} \in [-\pi, \pi]$, $i \in \{1, 2\}$, and $j \in \{1, 2, 3\}$.
Third, the number of spontaneous parametric down-conversion entangled photon pair denoted as $\mu$ is considered for the multiphoton effects.

Finally, two preprocessing methods are proposed in Ref.~\cite{Liu21} to enhance the experimental loss tolerance of undetected events; hence, the key rate is increased in the protocol. The probability of discarding key bits ``1" randomly and independently for Alice and Bob is denoted as $p\in[0, 1]$, which is for the random post-selection method in Ref.~\cite{Liu21} to remove a fraction of events of particles not detected by their detectors. At the same time, the probability of flipping each of Alice's remaining key bits to generate her noisy raw keys is denoted as $p_N\in[0, 1]$, which is for the noisy preprocessing method in Ref.~\cite{Liu21} to decrease the correlations between Alice's key and Eve's information. Then, a set of experimental parameters in the simulation of DIQKD~\cite{Liu21} is $\{\alpha, \phi_{Ai}, \phi_{Bj}, \mu, p, p_N\}$. By contrast, a set of experimental parameters in the simulation of EDIQKD is $\{\alpha, \mu, (p, p_N)\}$ (if two preprocessing methods are applied).

The required detection efficiency in our protocol is $88.7\%$ as shown in Fig.~\ref{keyratedetecteff}, which is $2.5\%$ lower than that of the usual DIQKD protocol $91.2\%$ \cite{Liu21}.
Furthermore, if we consider two preprocessing methods proposed in Ref.~\cite{Liu21}, i.e., random post-selection and noisy preprocessing, the required detection efficiency of our protocol releases from $88.7\%$ to $88.5\%$.
Given the improved fidelity of entangled state $99.8\%$, which can be achieved by the experimental system in Ref.~\cite{Liu21}, the required detection efficiency of our protocol decreases from $88.7\%$ to $88.2\%$.
With the efficiency values and errors of practical optical elements, as shown in Fig.~\ref{experimentsetup}(d), considered in Ref.~\cite{Liu21},
the required detection efficiencies of our protocol, $88.7\%$ and $88.2\%$, can be achieved, respectively.

%If the experimental data can not be used to build a completely positive process matrix, we use maximum-likelihood estimation (MLE) technique \cite{Brien20} to estimate the process matrix, and the threshold of detection efficiency will rise about $0.2\% \sim 0.4\%$. For the results shown in Fig.~\ref{keyratedetecteff} with MLE, the threshold of detection efficiency for key generation in our protocol becomes $88.9\%$ with the entangled state fidelity $99.52\%$ (plotted by blue curve), and the detection efficiency threshold drops to $88.5\%$ for the entangled state fidelity $100\%$ (plotted by orange curve). With random post-selection, noisy preprocessing and practical imperfections, the threshold of detection efficiency becomes $88.8\%$ (plotted by purple curve).

\begin{figure}[t]
\includegraphics[width=8cm]{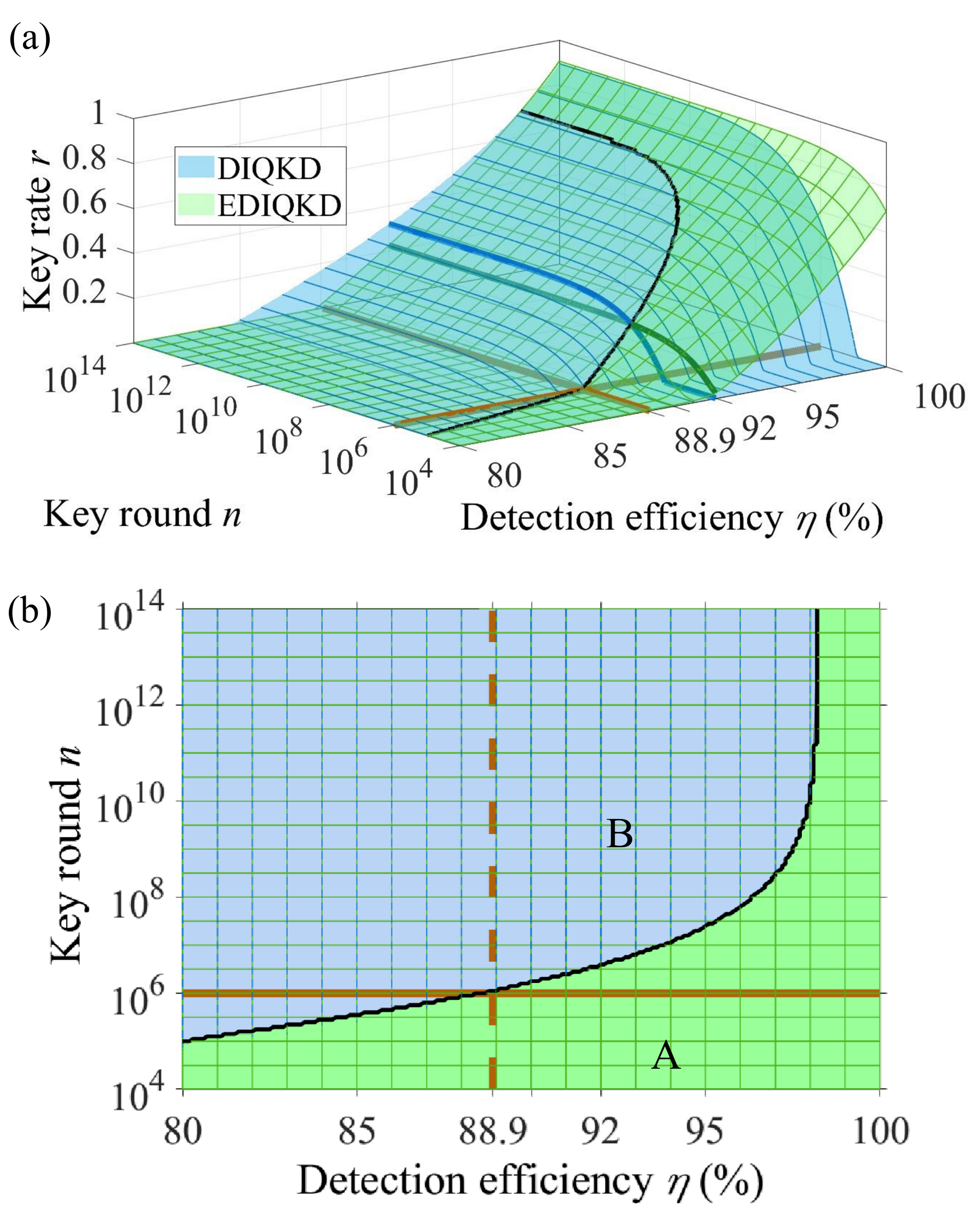}
\caption{Variation of key rates in EDIQKD and DIQKD protocols with numbers of key rounds under different detection efficiencies. (a) The key rate in the EDIQKD (DIQKD) protocol is plotted by the green (blue) transparent surface with grids (dashed lines). The black line is the intersection of two surfaces, and the brown dashed line is the required detection efficiency $88.9\%$ for the apparent difference between key rates in EDIQKD and DIQKD protocol with the number of key rounds less than $10^{6}$ (marked by the brown solid line). Under the detection efficiency $92\%$, the key rate in EDIQKD (DIQKD) protocol is plotted by the dark green (blue) line, and the efficiency of EDIQKD protocol is $10^{2.04} \sim 109.65$ times higher than that of DIQKD protocol. (b) A phase plot of comparison between key rates in EDIQKD and DIQKD protocols. Separated by the black line, the key rate in the EDIQKD (DIQKD) protocol is higher than that in the DIQKD (EDIQKD) protocol in area A (B).}\label{3D}
\end{figure}

To show how more efficient EDIQKD protocol than DIQKD protocol~\cite{Murta19} is to guarantee security for the imperfect detection efficiency, we consider the best-entangled state fidelity $99.8\%$ and execute the simulation similar to the former in Fig.~\ref{keyratedetecteff} (see Appendix~\ref{appendix:6} for details). The comparison between key rates in two protocols is shown in Figs.~\ref{3D}(a) and~\ref{3D}(b), and the efficiency factors $E_\text{f}$ for the varying detection efficiency are shown in Table~\ref{3D_b_table}. Without loss of generality, we choose $0.1\%$ for the precision. Under the detection efficiency $88.9\%$ ($88.8\%$), the efficiency of EDIQKD protocol is $10^{0.55} \sim 3.55 >1$ ($10^{-0.03} \sim 0.933 <1$) times that of DIQKD protocol, and thus we choose the detection efficiency $88.9\%$ for our following discussion. With the required detection efficiency $88.9\%$ and the number of key rounds less than $10^{6}$, the key rate in EDIQKD protocol is higher than that in DIQKD protocol, and the difference becomes more evident for the greater detection efficiency, which will be practically achieved in the future experiment. Under the detection efficiency $92\%$, the efficiency of EDIQKD protocol is $10^{2.04} \sim 109.65$ times higher than that of DIQKD protocol. It is worth noting that considering the best detection efficiency $89.73\%$ which can be achieved with experimental setup in Ref.~\cite{Liu21} [see Fig.~\ref{experimentsetup}(d) for details], the efficiency of EDIQKD protocol is $10^{1.44} \sim 27.54$ times higher than that of DIQKD protocol, as shown in Table~\ref{3D_b_table}.

\section{Conclusion and outlook}
We propose a prepare-and-measure QKD protocol and examine the protocol with the device-independent scenario. We show the proposed protocol, which security relies on ruling out any mimicry using the classical initial, transmission, and final state, is secure under collective and coherent attacks with IID assumption. It has a higher key rate than usual entanglement-based DIQKD since our protocol realizes key distribution and characterizes the feature of the channel at the same time. The efficiency in our protocol is two orders of magnitude higher than that in the DIQKD protocol. Given that the fidelity of the entangled state is $99.52\%$ ($99.8\%$), the required detection efficiency in our protocol is $88.7\%$ ($88.2\%$), which is achievable in the photonic experiment in Ref.~\cite{Liu21} and shows that our protocol is experimentally feasible.

Considering the best-entangled state fidelity $99.8\%$~\cite{Liu21} and the detection efficiency over $88.9\%$, the key rate in EDIQKD protocol is higher than that in DIQKD protocol with the number of key rounds less than $10^{6}$.
For example, when the detection efficiency is higher than and equal to $92\%$, the efficiency of the EDIQKD protocol is two orders of magnitude higher than that of the DIQKD protocol.
Furthermore, considering the best detection efficiency $89.73\%$~\cite{Liu21}, the efficiency of the EDIQKD protocol is $10^{1.44} \sim 27.54$ times higher than that of the DIQKD protocol.
The efficiency of the EDIQKD protocol will become more evident when greater detection efficiency is practically achieved in the future.
Our protocol remains more efficient under the imperfect detection efficiency $\eta \geq 88.9\%$.

Our EDIQKD protocol and the secure analyzing method provide new insight into DI quantum information. They may help realize quantum information tasks under different scenarios and purposes, such as DI quantum computing~\cite{NiNi}, quantum networks~\cite{Pirker20182018}, and one-sided DIQKD protocol~\cite{Branciard20122012}.

\section*{Acknowledgements}
We thank T.-Y. Tsai and Y.-C. Liang for helpful comments and discussions.
This work is partially supported by the National Science and Technology Council, Taiwan, under Grant Numbers NSTC 107-2628-M-006-001-MY4, NSTC 111-2119-M-007-007, and NSTC 111-2112-M-006-033.

\appendix
\renewcommand\appendixname{APPENDIX}

\section{GENUINELY CLASSICAL PROCESSES}\label{appendix:1}%Genuinely classical processes

We define a genuinely classical process (GCP) as a set of three steps describing the system state and its evolution \cite{Chen20}. In particular, the input system is considered to be a physical object with properties satisfying the assumption of realism \cite{Brunner14}. The system then evolves in accordance with classical stochastic theory \cite{Breuer&Petruccione02} and satisfies the classical dynamics. Finally, once the process is complete, the system resides in its final output state with properties satisfying the assumption of realism.

Since a GCP treats the initial system as a physical object with properties satisfying the assumption of realism, the system can be modeled as a state described by a fixed set of physical properties $\textbf{v}_{\xi}$. Assume that the system is described by three properties, say $V_{1}$, $V_{2}$ and $V_{3}$, where each property has two possible states. There therefore exist $2^{3}=8$ sets underlying the classical object: $\textbf{v}_{\xi}(\text{v}_{1},\text{v}_{2},\text{v}_{3})$, where $\text{v}_{1},\text{v}_{2},\text{v}_{3}\in\{+1,-1\}$ represent the possible measurement outcomes for $V_{1}$, $V_{2}$ and $V_{3}$, respectively. The subsequent classical evolution of the system changes the system from an initial state $\textbf{v}^{(A)}_{\xi}$ to a final state $\textbf{v}^{(B)}_{\mu}$ according to the transition probabilities $\Omega_{\xi\mu}$. Therefore, the relationship between a specific input state of the $i$th physical property of the system, e.g., $\text{v}_{i}=v^{(A)}_{i}$, and a specific output state of the $j$th physical property $v^{(B)}_{j}$ can be characterized by
\begin{equation}
P(v^{(B)}_{j}|v^{(A)}_{i})=\sum_{\xi,\mu}P(\textbf{v}^{(A)}_{\xi}|v^{(A)}_{i})\Omega_{\xi\mu}P(v^{(B)}_{j}|\textbf{v}^{(B)}_{\mu}),\label{resultingstate}
\end{equation}
where $v^{(A)}_{i}$ and $v^{(B)}_{j}$ correspond to the measurement outcomes $a_i$ and $b_j$, respectively, in Sec. \ref{section:2}.
Let PT be used to systematically exploit the experimentally measurable quantities given in Eq.~(\ref{resultingstate}). Suppose that the GCP can be completely characterized with a positive Hermitian matrix, called the process matrix, of the form
\begin{equation}
\chi_{\text{GC}}=  \frac{1}{2}\left[ \begin{matrix}
    \hat{I}_{gc}+\hat{V}_{gc,3} & \hat{V}_{gc,1}+i\hat{V}_{gc,2} \\
    \hat{V}_{gc,1}-i\hat{V}_{gc,2} & \hat{I}_{gc}-\hat{V}_{gc,3}
    \end{matrix}
\right],
\end{equation}
where $\hat{I}_{gc}\equiv\rho_{v^{(A)}_{i}=+1}+\rho_{v^{(A)}_{i}=-1}$ and $\hat{V}_{gc,i}\equiv\rho_{v^{(A)}_{i}=+1}-\rho_{v^{(A)}_{i}=-1}$ for $i=1,2,3$. Using Eq.~(\ref{resultingstate}) and state tomography \cite{Vogel89}, the density operator of the output system conditioned on a specific initial state $v^{(A)}_{i}$ is given by
\begin{equation}
\rho_{v^{(A)}_{i}}=\frac{1}{2}(\hat{I}+\sum_{j=1}^{3}\sum_{v^{(B)}_{j}=\pm 1}v^{(B)}_{j}P(v^{(B)}_{j}|v^{(A)}_{i})\hat{V}^{(B)}_{j}),\label{rhogc}
\end{equation}
where $\hat{I}$ denotes the identity operator and the observables $\hat{V}_{j}^{(B)}$ are the quantum analogues of the physical properties $V_{j}^{(B)}$ and are complementary to each other.

Suppose that, by using the PT procedure described above, a normalized process matrix $\chi_{\text{expt}}$ for the experimentally available data $P(v^{(B)}_{j}|v^{(A)}_{i})$ is created. Suppose further that the process fidelity of $\chi_{\text{expt}}$ and $\chi_{I}$, which is the process matrix of the identity unitary transformation, is used to evaluate the performance of the experimental process. For a given set of observables $\{\hat{V}_{i}^{(A)},\hat{V}^{(B)}_{j}|i,j=1,2,3\}$, if the process fidelity satisfies Eq.~(\ref{fidelity})
%\begin{equation}
%\red{F_{\text{expt}}\equiv \text{tr}(\chi_{\text{expt}}\chi_{I})> F_{\text{GC}}\equiv\max_{\chi_{\text{GC}}} \text{tr}(\chi_{\text{GC}}\chi_{I}),\label{fidelity2}}
%\end{equation}
then the experimental process with its process matrix $\chi_{\text{expt}}$ is qualified as truly nonclassical and is close to target process. The overriding goal of Eq.~(\ref{fidelity}) is to rule out the best classical mimicry of ideal transformation $\chi_{I}$. Such a capability of genuinely classical mimicry can be evaluated by performing the following mathematical maximization task via semi-definite programming with MATLAB \cite{Lofberg, sdpsolver}: $\max_{\chi_{\text{GC}}}\hspace{3pt}\text{tr}(\chi_{\text{GC}}\chi_{I})$,
such that $\chi_{\text{GC}}\geq 0,\hspace{3pt}\text{tr}(\chi_{\text{GC}})=1$, $\Omega_{\xi\mu} \geq 0$ $\forall\ \xi$, $\mu$. The above constraints ensure that the GCP matrix $\chi_{\text{GC}}$ satisfies both the definitions of process fidelity and a density operator. Specifically, as the observables for the PT procedure are chosen as $\hat{V}^{(A)}_{1}=X, \hat{V}^{(A)}_{2}=Y, \hat{V}^{(A)}_{3}=Z$ for the input states and $\hat{V}^{(B)}_{1}=UXU^{\dag}, \hat{V}^{(B)}_{2}=UYU^{\dag}, \hat{V}^{(B)}_{3}=UZU^{\dag}$ for the output states, where $U=\left|0\right\rangle\!\!\left\langle0\right|+\exp(i\pi/4)\left|1\right\rangle\!\!\left\langle1\right|$ and $X$, $Y$, and $Z$ are the Pauli matrices, the clearest distinction possible is obtained between the classical result and the quantum mechanical prediction. The closest similarity to identity process is
$F_{\text{GC}}\sim0.8536,$
under the above measurement setting. By analysing all the measure results $P(v^{(B)}_{j}|v^{(A)}_{i})$ and getting the process matrix $\chi_\text{expt}$, the participants, Alice and Bob, can use the fidelity criteria to make sure whether the channel between them is faithful.

\section{Comparison of secure key rate in the asymptotic limit of infinite key lengths}\label{appendix:2}

To compare with the entangled DIQKD protocol, we first show how to implement our EDIQKD protocol through entangled pairs, which Alice prepares. With entangled states, Alice prepares a specific state by performing corresponding projective measurements on one particle of her entangled pair. The projectors used are transposed from the input states. The process matrix considered in Eq.~(\ref{fidelity}) can be obtained by collecting Alice's and Bob's measurement results. Through the criterion Eq.~(\ref{fidelity}), Alice and Bob can rule out any classical mimicry of the local realistic model.

\begin{figure}[t]
\includegraphics[width=7.9cm]{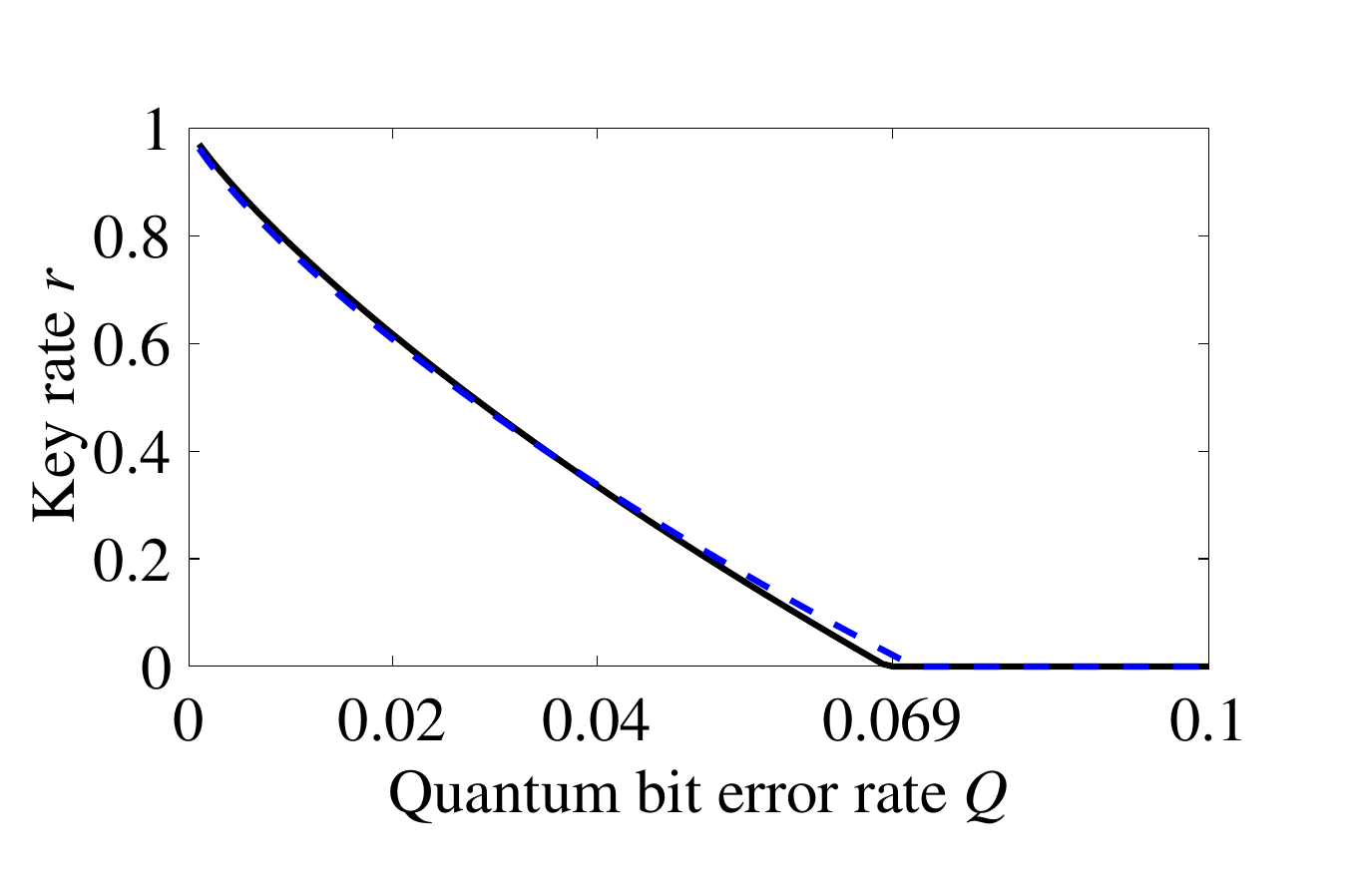}
\caption{Extractable secret-key rate against collective attacks in the EDIQKD protocol (black line) and the usual DIQKD protocol (blue-dashed line).
For the EDIQKD (DIQKD) protocol, the key rate $r=0$ while the QBER $Q = 6.9\%$ ($7.1\%$).
}\label{keyrate}
\end{figure}

For the cloner \cite{Scarani05,Chiu16} considered in Sec. \ref{section:3_1} that $\lambda_{00}=(1-p)$ and $\lambda_{01}=\lambda_{10}=\lambda_{11}=p/3$ since the eavesdropping attacks by using a cloning machine on three orthogonal bases are the same, the key rate is plotted as a function of quantum bit error rate (QBER) $Q$ in Fig.~\ref{keyrate}, where $Q=2p/3$. While key rate $r=0$, the QBER $Q=6.9\%$ and the process fidelity $F=0.8656>F_{\text{GC}}=0.8536$.
We compare our key rate with the key rate under the usual DIQKD protocol \cite{Acin06} against collective attacks.
%In the usual DIQKD protocol, Alice and Bob first share a quantum channel consisting of a source of entangled particles, and then they use Clauser-Horne-Shimony-Holt (CHSH) inequality \cite{Clauser69} to examine the entangled sources. If the entangled pairs are faithful, the raw key is extracted from measurement results of specific basis.
%Considering collective attack, the critical QBER $Q$ is $7.1\%$ \cite{Acin07} (as shown in Fig.~\ref{keyrate}, where the number of key rounds $n = 10^{15}$ is considered).
The critical QBER $Q$ in the usual DIQKD protocol is $7.1\%$ \cite{Acin07}, as shown in Fig.~\ref{keyrate}, where the number of key rounds $n = 10^{15}$ is considered.
%The critical QBER $Q$ in EDIQKD protocol is stricter than but close to that in the usual DIQKD protocol, which shows that our protocol is comparable in the asymptotic limit of infinite key lengths (for $n \geq 10^{15}$ \cite{Rotem2018}).
Therefore, our protocol is comparable in the asymptotic limit of infinite key lengths (for $n \geq 10^{15}$ \cite{Rotem2018}).

\section{CALCULATION FOR TRACE DISTANCE FOR SECRECY}\label{trD}

To calculate the secrecy for key rate, we start with an universal quantum copying machine (UQCM) \cite{Masab22} acting on the tripartite composite system among Alice, Bob, and Eve.
The process of UQCM, $\mathcal{E}_{\text{UQCM}}$, is a 3-qubit process represented by the quantum circuit in Ref.~\cite{Masab22}, and the output cloned states are universal to the input state.
The input states are Alice's prepared state $\rho_{A'|A}$, Eve's initial state $\ket{0}_{EE}\!\bra{0}$, and Eve's initial ancilla qubit $\ket{0}_{E'E'}\!\bra{0}$.
The output states are Bob's received state $\rho_{B|AE}$, Eve's copied system $E$, and Eve's received ancilla system $E'$.
Two cloned states of $\rho_{A'|A}$ with state fidelity $5/6 \sim 83.33\%$ are $\rho_{B|AE}$ and system $E$, and ancilla system $E'$ does not contain any information about $\rho_{A'|A}$.

Here, we will characterize the 2-qubit process $\mathcal{E}_{\text{ABE}}$ under the eavesdropper's attack by the process matrix $\chi_{\text{ABE}}$.
Considering a cloner that $\lambda_{00}=(1-p)$ and $\lambda_{01}=\lambda_{10}=\lambda_{11}=p/3$ since the eavesdropping attacks by using UQCM on three orthogonal bases are the same, then $\rho_{B|AE}$ can be represented as
\begin{eqnarray}
\rho_{B|AE} &=& p'[(1-\frac{2p}{3})\rho_{A'|A}+\frac{2p}{3}\rho^{\perp}_{A'|A}]\nonumber\\
&&+(1-p')\rho_{A'|A}\nonumber\\
&=&(1-\frac{2pp'}{3})\rho_{A'|A}+\frac{2pp'}{3}\rho^{\perp}_{A'|A},\nonumber\\
\end{eqnarray}
where $p'$ is the probability of Eve using UQCM, and $\rho^{\perp}_{A'|A}$ is the state orthogonal to $\rho_{A'|A}$.
Thus the QBER $Q$ is $2pp'/3$.
The state fidelity of the cloned state of $\rho_{A'|A}$ is $5/6$, so $1-2p/3 = 5/6$ and $p = 1/4$.
Then the relation between $p'$ and $Q$ is $p'=6Q$, which is the reason why the trace distance $D$ is a function of QBER $Q$.
The 2-qubit process of UQCM, $\mathcal{E}'_{\text{UQCM}}$, is the process for the input states, $\rho_{A'|A}$ and $\ket{0}_{EE}\!\bra{0}$, and the output cloned states, $\rho'_{BE}$.
The operator-sum representation~\cite{Nielsen&Chuang00} of $\mathcal{E}'_{\text{UQCM}}$ is derived from $\mathcal{E}_{\text{UQCM}}$, and hence we can obtain the output states $\rho'_{BE}$ corresponding to Alice's input states $\rho_{A'|A} \in \{\ketbra{0}{0}, \ketbra{1}{1}, \ketbra{+}{+}, \ketbra{R}{R}\}$.
Then 2-qubit output states $\rho_{BE}$ can be represented as
\begin{eqnarray}
\rho_{BE} = p'\rho'_{BE}+(1-p')(\rho_{A'|A}\otimes\ket{0}_{EE}\!\bra{0}).
\end{eqnarray}
To characterize the process $\mathcal{E}_{\text{ABE}}$, we use 1-qubit PT algorithm to derive the process matrix describing $\rho_{A'|A}\otimes\ket{0}_{EE}\!\bra{0}\rightarrow\rho_{A'|A}\otimes\ket{0}_{EE}\!\bra{0}$ and 2-qubit PT with the 2-qubit states $\rho\blue{'}_{BE}$ (or can be derivied from $\mathcal{E}_{\text{UQCM}}$ as mentioned above) to obtain $\chi_{\text{ABE}}$.

\begin{figure}[t]
\includegraphics[width=7.9cm]{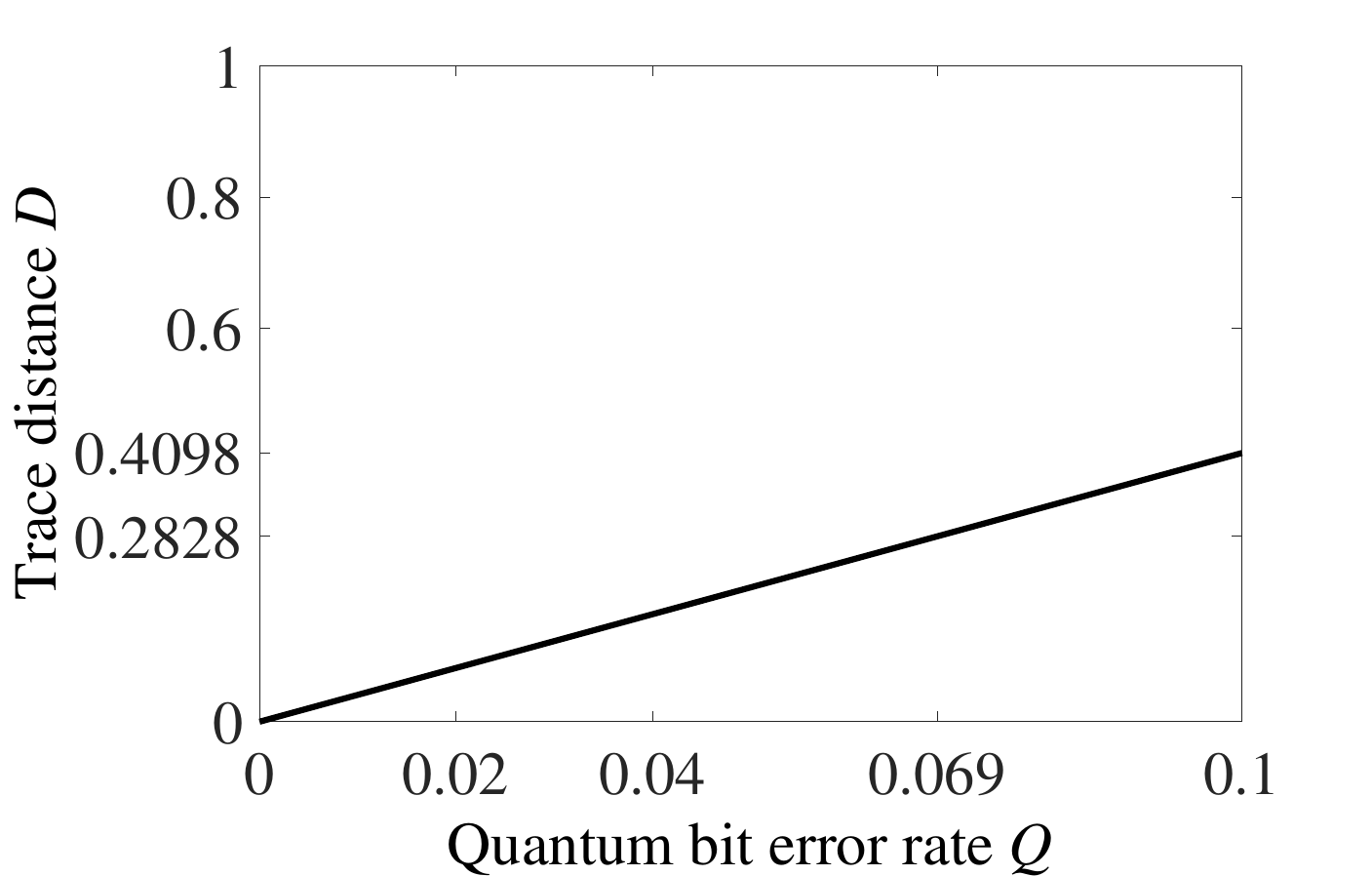}
\caption{The trace distance $D$ of the process matrix in the first phase of EDIQKD protocol under collective attacks to an ideal one for the secrecy versus the QBER $Q$. To guarantee the positive key rate ($r>0$) in EDIQKD protocol as $Q < 0.069$, $D$ should be smaller than $0.2828$.}
\label{tracedistance}
\end{figure}

The trace distance \cite{Nielsen&Chuang00} between two process matrix, $\chi_{1}$ and $\chi_{2}$, is defined by
\begin{eqnarray}
D(\chi_{1},\chi_{2}) = \frac{1}{2}\text{tr}|\chi_{1}-\chi_{2}|,
\end{eqnarray}
where $|\chi_{1}-\chi_{2}| = \sqrt{(\chi_{1}-\chi_{2})^{\dag}(\chi_{1}-\chi_{2})}$ is the positive square root of $(\chi_{1}-\chi_{2})^{\dag}(\chi_{1}-\chi_{2})$.
The trace distance $D$ of the process matrix in the protocol under collective attacks to an ideal one is plotted as a function of QBER $Q$ in Fig.~\ref{tracedistance}.

\section{DERIVATION OF KEY RATE}\label{appendix:4}

Here, we follow the proof in Ref.~\cite{Murta19} to present the proof of Eq.~(\ref{keyratef1}) and Eq.~(\ref{keyratef2}).
We assumed that the attack of the eavesdropper, Eve, is restricted to collective attacks.
According to Eq.~(B.2) in Ref.~\cite{Murta19} for the correctness in the second phase of the protocol, provided that the error correction protocol is successful, the QBER $Q$ of the $l'$-bit key is smaller than and equal to the error probability of the error correction protocol ${\epsilon}_\text{EC}$, i.e.,
\begin{eqnarray}
Q&\leq& \epsilon_\text{EC}.
\end{eqnarray}
For the secrecy in the second phase of the protocol, we need to estimate how far the process matrix in the protocol under collective attacks is from an ideal one.
Here, the processes that transform Alice's $n$-bit raw key into her $l'$-bit key before privacy amplification are considered.
According to Ref.~\cite{Tomamichel17}, we derive the supremum of the trace distance $D$ as following:
\begin{eqnarray}\label{hyjfvyuj}
D(\chi_{\text{ABE}},\chi_{\text{AB}}\otimes\chi_{\text{AE}})&\leq& 2^{-[H^{\epsilon_\text{s}}_\text{min}(A^n|EO_\text{EC})-l']/2} \nonumber\\&&+2\epsilon_\text{s},
\end{eqnarray}
where $H^{\epsilon_\text{s}}_\text{min}(A^n|EO_\text{EC})$ is the smooth min-entropy of $n$-bit Alice's raw key $A^n$ conditioned on Eve's information $E$ and exchanged information for error correction $O_\text{EC}$, $\epsilon_\text{s}$ is the smoothing parameter, and $l'$ is the length of a secret key before privacy amplification.

The following is the detailed derivation of Eq.~(\ref{hyjfvyuj}).
According to Eq.~(13) in Ref.~\cite{Tomamichel17} for the definition of the min-entropy, there exists a normalized state $\rho_E$ such that $\rho_{BE} \leq 2^{-\lambda}\text{id}_B\otimes \rho_E$, where $\lambda$ is the real number of the min-entropy, and $\text{id}_B$ is the identity operator for Bob's unnormalized fully mixed state.
The fully mixed state represents a horizontally or vertically polarized photon with equal probabilities in a two-level system for example, and it also symbolizes a perfect key which is independent of the eavesdropper's system and uniformly distributed for the most randomness.
Considering the state sent from Alice for key rounds, $\rho_{A'|A} =\ket{i}_{AA}\!\bra{i}$, $i \in \{0,1\}$, we have
%\begin{eqnarray}
%&&\red{\frac{1}{2}\!\sum_{i=0}^{1}\!\ket{i}_{\!AA}\!\!\bra{i}\!\otimes\!\rho_{BE}}\!\!\nonumber\\
%&\leq&\!\! \frac{1}{2}\!\sum_{i=0}^{1}\!\ket{i}_{\!AA}\!\!\bra{i}\!\otimes\!(2^{-\lambda}\text{id}_B\!\otimes\! \rho_E)\nonumber\\
%&=&\!\!2^{-\lambda}\red{[(\frac{1}{2}\!\sum_{i=0}^{1}\!\ket{i}_{\!AA}\!\!\bra{i}\!\otimes\! \text{id}_B) \!\otimes\!(\frac{1}{2}\!\sum_{i=0}^{1}\!\ket{i}_{\!AA}\!\!\bra{i}\!\otimes\! \rho_E)]},\nonumber\\
%\end{eqnarray}
%and thus
\begin{eqnarray}
&\frac{1}{2}&\!\sum_{i=0}^{1}\!\mathcal{E}_{\text{ABE}}(\ket{i}_{AA}\!\bra{i}\!\otimes\!\ket{0}_{EE}\!\bra{0})\nonumber\\
&&\leq2^{-\lambda}\text{id}_B\!\otimes\! \rho_E=\mathcal{E}_{\text{Dep}}\!\otimes\!\mathcal{E}_{\text{AE}}(\ket{i}_{AA}\!\bra{i}\!\otimes\!\ket{0}_{EE}\!\bra{0}),\nonumber\\
\end{eqnarray}
and
\begin{eqnarray}
\chi_{\text{ABE}}&\leq&2^{-\lambda}\hat{I}_{\text{AB}}'\otimes \chi_{\text{AE}},
\end{eqnarray}
where $\hat{I}_{\text{AB}}'$ is the identity operator for the unnormalized fully depolarizing process $\mathcal{E}_{\text{Dep}}$.
The fully depolarizing process symbolizes that the eavesdropper can't determine what the process between Alice and Bob is without performing attacks.

\textbf{Definition (Min-entropy).} For any sub-normalized process of generating $l'$-bit key in $n$ key rounds of EDIQKD protocol, $\chi_{\text{ABE}} \in S'(ABE)$, we define min-entropy of $n$-bit Alice's raw key $A^n$ conditioned on Eve's information $E$ and exchanged information for error correction $O_\text{EC}$ as
$H_\text{min}(A^n|EO_\text{EC}) = \text{sup}\{ \lambda \in \mathbb{R}: \exists\ \chi_{\text{AE}} \in S(AE)\ \text{such that } \chi_{\text{ABE}} \leq 2^{-\lambda}\hat{I}_{\text{AB}}\otimes \chi_{\text{AE}}\}$,
where $S'(ABE)$ denotes the set of positive semi-definite process matrices with trace less than one (called sub-normalized) on systems A, B, and E, ``\text{sup}'' means supremum, $S(AE)$ denotes the set of normalized processes on systems A and E, and $\hat{I}_{\text{AB}} = (\hat{I}_{\text{AB}}')^{\otimes l'}$.

By definition of the min-entropy, there exists a process matrix $\chi_{\text{AE}} \in S(AE)$ such that
\begin{eqnarray}
\chi_{\text{ABE}} \leq 2^{-H_\text{min}(A^n|EO_\text{EC})}\hat{I}_{\text{AB}}\otimes \chi_{\text{AE}}.
\label{S4}
\end{eqnarray}
Next, we represent the trace distance as
\begin{eqnarray}
D(\chi_{\text{ABE}},\chi_{\text{AB}}\otimes\chi_{\text{AE}})
\!&=&\!\frac{1}{2}||\chi_{\text{ABE}}-\chi_{\text{AB}}\otimes\chi_{\text{AE}}||_1,\nonumber\\
\label{S5}
\end{eqnarray}
where $||\cdot||_1$ denotes the Schatten 1-norm, and $\chi_{\text{AB}} = [(1/2)\sum_{i=0}^{1}\ket{i}_{AA}\!\bra{i}\otimes (1/2)\text{id}_B]^{\otimes l'} = (1/2^{l'})\hat{I}_{\text{AB}}$ is the normalized fully depolarizing process.

By applying H$\ddot{\text{o}}$lder's inequality for Schatten norms~\cite{Bhatia97}, we obtain
\begin{eqnarray}
&||&\chi_{\text{ABE}}-\chi_{\text{AB}}\otimes\chi_{\text{AE}}||_1^2\nonumber\\
&\leq&\!\! |\!|\hat{I}_{\text{AB}}\!\!\otimes\!\chi_{\text{AE}}^{1\!/\!2}\!|\!|_2^2
|\!|\hat{I}_{\text{AB}}\!\!\otimes\!\chi_{\text{AE}}^{-\!1\!/\!2}\!( \chi_{\text{ABE}}\!-\!\chi_{\text{AB}}\!\!\otimes\!\chi_{\text{AE}})|\!|_2^2\nonumber\\
&=&\!\!2^{l'}\!\text{tr}(\chi_{\text{AE}})
\text{tr}[\hat{I}_{\text{AB}}\!\!\otimes\!\chi_{\text{AE}}^{-1}
\!(\chi_{\text{ABE}}\!-\!\chi_{\text{AB}}\!\!\otimes\!\chi_{\text{AE}})^2]\nonumber\\
&=&\!\!2^{l'}\!\text{tr}[(\hat{I}_{\text{AB}}\!\!\otimes\!\chi_{\text{AE}}^{-1}\!)\chi_{\text{ABE}}^2]\nonumber\\
&&-2^{l'}\!\text{tr}[(\hat{I}_{\text{AB}}\!\otimes\!\chi_{\text{AE}}^{-1})
(\frac{1}{2^{l'}}\hat{I}_{\text{AB}}\!\otimes\!\chi_{\text{AE}})^2]\nonumber\\
&=&\!\!2^{l'}\!\text{tr}[(\hat{I}_{\text{AB}}\!\!\otimes\!\chi_{\text{AE}}^{-1}\!)\chi_{\text{ABE}}^2]
-\text{tr}(\chi_{\text{AE}}^{-1}\chi_{\text{AE}}^2)\nonumber\\
&=&\!\!2^{l'}\!\text{tr}[(\hat{I}_{\text{AB}}\!\!\otimes\!\chi_{\text{AE}}^{-1}\!)\chi_{\text{ABE}}^2]-1,
\label{S6}
\end{eqnarray}
where $||\cdot||_2$ denotes the Schatten 2-norm, and $||T||_2 = [\text{tr}(T^{\dag}T)]^{1/2}$ for a matrix $T$ and $\text{tr}(\hat{I}_{\text{AB}}) = 2^{l'}$ are used to arrive at the first and the third equality. Substituting Eq.~(\ref{S6}) in Eq.~(\ref{S5}), we have
\begin{eqnarray}
&D&(\chi_{\text{ABE}},\chi_{\text{AB}}\otimes\chi_{\text{AE}})\nonumber\\
&\leq&\frac{1}{2}\sqrt{2^{l'}\text{tr}[(\hat{I}_{\text{AB}}\otimes\chi_{\text{AE}}^{-1})\chi_{\text{ABE}}^2]-1}\nonumber\\
&\leq&\frac{1}{2}\sqrt{2^{l'}\text{tr}[(\hat{I}_{\text{AB}}\otimes\chi_{\text{AE}}^{-1})\chi_{\text{ABE}}^2]}.
\end{eqnarray}
By Eq.~(\ref{S4}) and the operator anti-monotonicity of the inverse, we have
$
\hat{I}_{\text{AB}}\otimes \chi_{\text{AE}}^{-1}
\leq
2^{-H_\text{min}(A^n|EO_\text{EC})}
\chi_{\text{ABE}}^{-1}
$, and then we obtain
\begin{eqnarray}
&D&(\chi_{\text{ABE}},\chi_{\text{AB}}\otimes\chi_{\text{AE}})\nonumber\\
&\leq&\frac{1}{2}\sqrt{2^{l'}\text{tr}[(2^{-H_\text{min}(A^n|EO_\text{EC})}
\chi_{\text{ABE}}^{-1})\chi_{\text{ABE}}^2]}\nonumber\\
&=&\frac{1}{2}\sqrt{\text{tr}(\chi_{\text{ABE}})}2^{-[H_\text{min}(A^n|EO_\text{EC})-l']/2}.
\end{eqnarray}

Let $\tilde{\chi}_{\text{ABE}} \in S'(ABE)$ be a sub-normalized process matrix such that $H^{\epsilon_\text{s}}_\text{min}(A^n|EO_\text{EC})$ for $\chi_{\text{ABE}}$ is equal to $H_\text{min}(A^n|EO_\text{EC})$ for $\tilde{\chi}_{\text{ABE}}$ and $D(\tilde{\chi}_{\text{ABE}},\chi_{\text{ABE}})\leq \epsilon_\text{s}$,
where $H^{\epsilon_\text{s}}_\text{min}(A^n|EO_\text{EC})$ is the smooth min-entropy of $n$-bit Alice's raw key $A^n$ conditioned on Eve's information $E$ and exchanged information for error correction $O_\text{EC}$,
and $\epsilon_\text{s}$ is the smoothing parameter.
Then with the equation $\text{tr}(\tilde{\chi}_{\text{ABE}}) \leq 1$ we obtain
\begin{eqnarray}
&D&(\tilde{\chi}_{\text{ABE}},\chi_{\text{AB}}\otimes\tilde{\chi}_{\text{AE}})\nonumber\\
&\leq&\frac{1}{2}\sqrt{\text{tr}(\tilde{\chi}_{\text{ABE}})}2^{-[H_\text{min}(A^n|EO_\text{EC})-l']/2}\nonumber\\
&\leq&2^{-[H^{\epsilon_\text{s}}_\text{min}(A^n|EO_\text{EC})-l']/2}.
\end{eqnarray}
Using the monotonicity of the purified distance under completely positive and trace-preserving maps, we have $D(\tilde{\chi}_{\text{AE}},\chi_{\text{AE}})\leq D(\tilde{\chi}_{\text{ABE}},\chi_{\text{ABE}})\leq \epsilon_\text{s}$. Finally, by using the triangle inequality for the trace distance~\cite{Tomamichel16}, we obtain\\
\begin{eqnarray}
&D&(\chi_{\text{ABE}},\chi_{\text{AB}}\otimes\chi_{\text{AE}})\nonumber\\
&\leq&D(\tilde{\chi}_{\text{ABE}},\chi_{\text{ABE}})+D(\tilde{\chi}_{\text{AE}},\chi_{\text{AE}})\nonumber\\
&&+D(\tilde{\chi}_{\text{ABE}},\chi_{\text{AB}}\otimes\tilde{\chi}_{\text{AE}})\nonumber\\
&\leq&2\epsilon_\text{s}+2^{-[H^{\epsilon_\text{s}}_\text{min}(A^n|EO_\text{EC})-l']/2}.
\end{eqnarray}

Thus the secret key length $l'$ can be estimated by $H^{\epsilon_\text{s}}_\text{min}(A^n|EO_\text{EC})$, i.e.,
\begin{eqnarray}
l' = H^{\epsilon_\text{s}}_\text{min}(A^n|EO_\text{EC}).
\end{eqnarray}
According to Eqs.~(23) and~(24) in Ref.~\cite{Murta19}, the secret key length $l$ after privacy amplification is bounded by
\begin{eqnarray}\label{eqs10}
l &\geq& H^{\epsilon_\text{s}}_\text{min}(A^n|EO_\text{EC})-2\text{log}_2\left(\frac{1}{2\epsilon_\text{PA}}\right),
\end{eqnarray}
where $\epsilon_\text{PA}$ is the error probability of the privacy amplification protocol.
To consider the leakage in the error correction protocol by Eq.~(A.16) in Ref.~\cite{Murta19} for the smooth min-entropy conditioned on classical information, we obtain
\begin{eqnarray}\label{eqs11}
H^{\epsilon_\text{s}}_\text{min}\!(\!A^n|EO_\text{EC}) \!\!&\geq&\!\! H^{\epsilon_\text{s}}_\text{min}\!(\!A^n|E)\! -\! \text{leak}_{\text{EC}},
\end{eqnarray}
where $H^{\epsilon_\text{s}}_\text{min}(A^n|E)$ is the smooth min-entropy of $n$-bit Alice's raw key $A^n$ conditioned on Eve's information $E$, and $\text{leak}_{\text{EC}}$ denotes the minimum classical information sent from Alice to Bob for the error correction.

Here, we will estimate the lower bound of $H^{\epsilon_\text{s}}_\text{min}(A^n|E)$. By using Eqs.~(25),~(B.15), and~(B.16) in Ref.~\cite{Murta19}, we can decompose $H^{\epsilon_\text{s}}_\text{min}(A^n|E)$ into the sum of the conditional entropy of a single round,
\begin{eqnarray}\label{eqs12}
H^{\epsilon_\text{s}}_\text{min}(A^n|E) &\geq& n H(A|E) \nonumber\\&&- \sqrt{n}\!\!\left(\!\!4\text{log}_2(2\sqrt{2}\!+\!1)\sqrt{\text{log}_2\frac{2}{\epsilon_\text{s}^2}}\ \right),\nonumber\\
\end{eqnarray}
where $H(A|E)$ is the von-Neumann entropy of Alice's outcome $A$ conditioned on Eve's information $E$ in one round.
Then we can estimate the lower bound of $H(A|E)$ by that the mutual information $I(A:E)$ is constrained by the Holevo bound \cite{Chiu16} [Eq.~(\ref{eq4})]:
\begin{equation}
I(A:E) \leq S(\rho_{EE'})-\frac{1}{2}\sum_{a_i=\pm1}S(\rho_{EE'|a_i}).
\end{equation}
We assume Alice's outcome $A$ is so random that the von-Neumann entropy of Alice's outcome $H(A)$ is equal to one, then we have
\begin{eqnarray}\label{eqs14}
H(A|E) &=& H(A) - I(A:E) \nonumber \\
&\geq& \!1\!-\!\![S(\rho_{EE'})-\!\frac{1}{2}\!\sum_{a_i=\pm1}\!\!S(\rho_{EE'|a_i})]. \nonumber \\
\end{eqnarray}
Substituting Eq.~(\ref{eqs14}) in Eq.~(\ref{eqs12}), we obtain the lower bound of $H^{\epsilon_\text{s}}_\text{min}(A^n|E)$,
\begin{eqnarray}\label{eqs15}
H^{\epsilon_\text{s}}_\text{min}\!(A^n|E) \!\!&\geq&\!\!
n\{1\!-\!\![S(\rho_{EE'})-\!\frac{1}{2}\!\!\!\sum_{a_i=\pm1}\!\!\!\!S(\rho_{EE'|a_i})]\} \nonumber\\
&&- \sqrt{n}\!\!\left(\!\!4\text{log}_2(2\sqrt{2}\!+\!1)\sqrt{\text{log}_2\frac{2}{\epsilon_\text{s}^2}}\ \right).\nonumber\\
\end{eqnarray}

Here, we will estimate the upper bound of $\text{leak}_{\text{EC}}$. Given that the probability of aborting error correction protocol is no more than ${\epsilon}_\text{EC} + {\epsilon}_\text{EC}'$, we acquire the upper bound of $\text{leak}_{\text{EC}}$ by using Eqs.~(B.10) and~(B.11) in Ref.~\cite{Murta19},
\begin{eqnarray}\label{eqs16}
\text{leak}_{\text{EC}} &\leq& H^{{{\epsilon}_\text{EC}'}/2}_\text{max}\!(\!A^n|B^n)\nonumber\\
&&+\text{log}_2\!\left(\!\frac{8}{{\epsilon}_\text{EC}'^2}+\frac{2}{2-{\epsilon}_\text{EC}'}\!\right)\!\!+\!\text{log}_2\!\left(\!\frac{1}{{\epsilon}_\text{EC}}\!\right),\nonumber\\
\end{eqnarray}
where $H^{{{\epsilon}_\text{EC}'}/2}_\text{max}(A^n|B^n)$ is the smooth max-entropy of $n$-bit Alice's raw key $A^n$ conditioned on $n$-bit Bob's raw key $B^n$. By using Eqs.~(26),~(B.15), and~(B.16) in Ref.~\cite{Murta19}, we can break $H^{{{\epsilon}_\text{EC}'}/2}_\text{max}(A^n|B^n)$ into the sum of the conditional entropy of a single round,
\begin{eqnarray}\label{eqs17}
H^{{{\epsilon}_\text{EC}'}/2}_\text{max}(A^n|B^n) \!\!&\leq&\!\! n H(A|B) \nonumber\\
&&\!\!+ \sqrt{n}\!\!\left(\!\!4\text{log}_2\!(2\sqrt{2}\!+\!\!1)\!\sqrt{\!\text{log}_2\frac{8}{{\epsilon}_\text{EC}'^2}}\right)\!\!,\nonumber\\
\end{eqnarray}
where $H(A|B)$ is the von-Neumann entropy of Alice's outcome $A$ conditioned on Bob's outcome $B$ in one round. $H(A|B)$ in key rounds is the binary entropy of QBER, i.e., $h(Q)$ \cite{Murta19}.
The mutual information $I(A:B)$ in test rounds is $1-h(F_{\text{expt}})$, and $H(A|B) = H(A) - I(A:B) = h(F_{\text{expt}})$ in test rounds. Then we can estimate $H(A|B)$ in one round by using Eq.~(A.8) in Ref.~\cite{Murta19},
\begin{eqnarray}\label{eqs18}
H(A|B) = (1-\gamma)h(Q)+\gamma h(F_{\text{expt}}),
\end{eqnarray}
where $\gamma$ is the fraction of test rounds.
Substituting Eqs.~(\ref{eqs17}) and~(\ref{eqs18}) in Eq.~(\ref{eqs16}), we obtain the upper bound of $\text{leak}_{\text{EC}}$,
\begin{eqnarray}\label{eqs19}
\text{leak}_{\text{EC}} &\leq&
n [(1-\gamma)h(Q)+\gamma h(F_{\text{expt}})] \nonumber\\
&&+ \sqrt{n}\!\!\left(\!\!4\text{log}_2(2\sqrt{2}\!+\!1)\sqrt{\text{log}_2\frac{8}{{\epsilon}_\text{EC}'^2}}\ \right)\nonumber\\
&&+\text{log}_2\!\left(\!\frac{8}{{\epsilon}_\text{EC}'^2}+\frac{2}{2-{\epsilon}_\text{EC}'}\!\right)\!\!+\!\text{log}_2\!\left(\!\frac{1}{{\epsilon}_\text{EC}}\!\right),\nonumber\\
\end{eqnarray}

Finally, substituting Eqs.~(\ref{eqs11}),~(\ref{eqs15}) and~(\ref{eqs19}) in Eq.~(\ref{eqs10}), we obtain the secret key length $l$,
\begin{eqnarray}\label{eqs20}
l &\geq&
n \{1-(1-\gamma)h(Q)-\gamma h(F_{\text{expt}})\nonumber\\
&&-[S(\rho_{EE'})-\frac{1}{2}\sum_{a_i=\pm1}S(\rho_{EE'|a_i})]\} \nonumber\\
&&- \sqrt{n}\!\!\left[\!4\text{log}_2\!(2\sqrt{2}\!+\!\!1)\!\!\left(\!\!\sqrt{\text{log}_2\frac{2}{\epsilon_\text{s}^2}} +\!\! \sqrt{\text{log}_2\frac{8}{{\epsilon}_\text{EC}'^2}} \right)\!\!\right]\nonumber\\
&&-\text{log}_2\!\left(\!\frac{8}{{\epsilon}_\text{EC}'^2}+\frac{2}{2-{\epsilon}_\text{EC}'}\!\right)\!\!-\!\text{log}_2\!\left(\!\frac{1}{{\epsilon}_\text{EC}}\!\right)\nonumber\\
&&-2\text{log}_2\left(\frac{1}{2\epsilon_\text{PA}}\right).
\end{eqnarray}
By using Eq.~(5) in Ref.~\cite{Murta19}, we obtain the secret key rate $r$ of an $l$-bit correct-and-secret key in $n$ key rounds of EDIQKD protocol,
\begin{eqnarray}\label{eqs21}
r &=& \frac{l}{n} \nonumber\\
&\geq& 1-(1-\gamma)h(Q)-\gamma h(F_{\text{expt}})\nonumber\\
&&-[S(\rho_{EE'})-\frac{1}{2}\sum_{a_i=\pm1}S(\rho_{EE'|a_i})] \nonumber\\
&&- \!\!\left\{\!\!\sqrt{n}\!\!\left[\!4\text{log}_2(\!2\sqrt{2}\!+\!\!1\!)\!\!\left(\!\!\sqrt{\text{log}_2\frac{2}{\epsilon_\text{s}^2}} \!\!+\!\! \sqrt{\text{log}_2\frac{8}{{\epsilon}_\text{EC}'^2}} \right)\!\!\right]\right.\nonumber\\
&&+\text{log}_2\!\left(\!\frac{8}{{\epsilon}_\text{EC}'^2}+\frac{2}{2-{\epsilon}_\text{EC}'}\!\right)\!\!+\!\text{log}_2\!\left(\!\frac{1}{{\epsilon}_\text{EC}}\!\right)\nonumber\\
&&\left.+2\text{log}_2\left(\frac{1}{2\epsilon_\text{PA}}\right)\right\}\frac{1}{n}.
\end{eqnarray}

\section{EXPECTED PHOTONIC EXPERIMENTAL SETUP}\label{appendix:5}

\begin{figure*}[t]
\includegraphics[width=16.5cm]{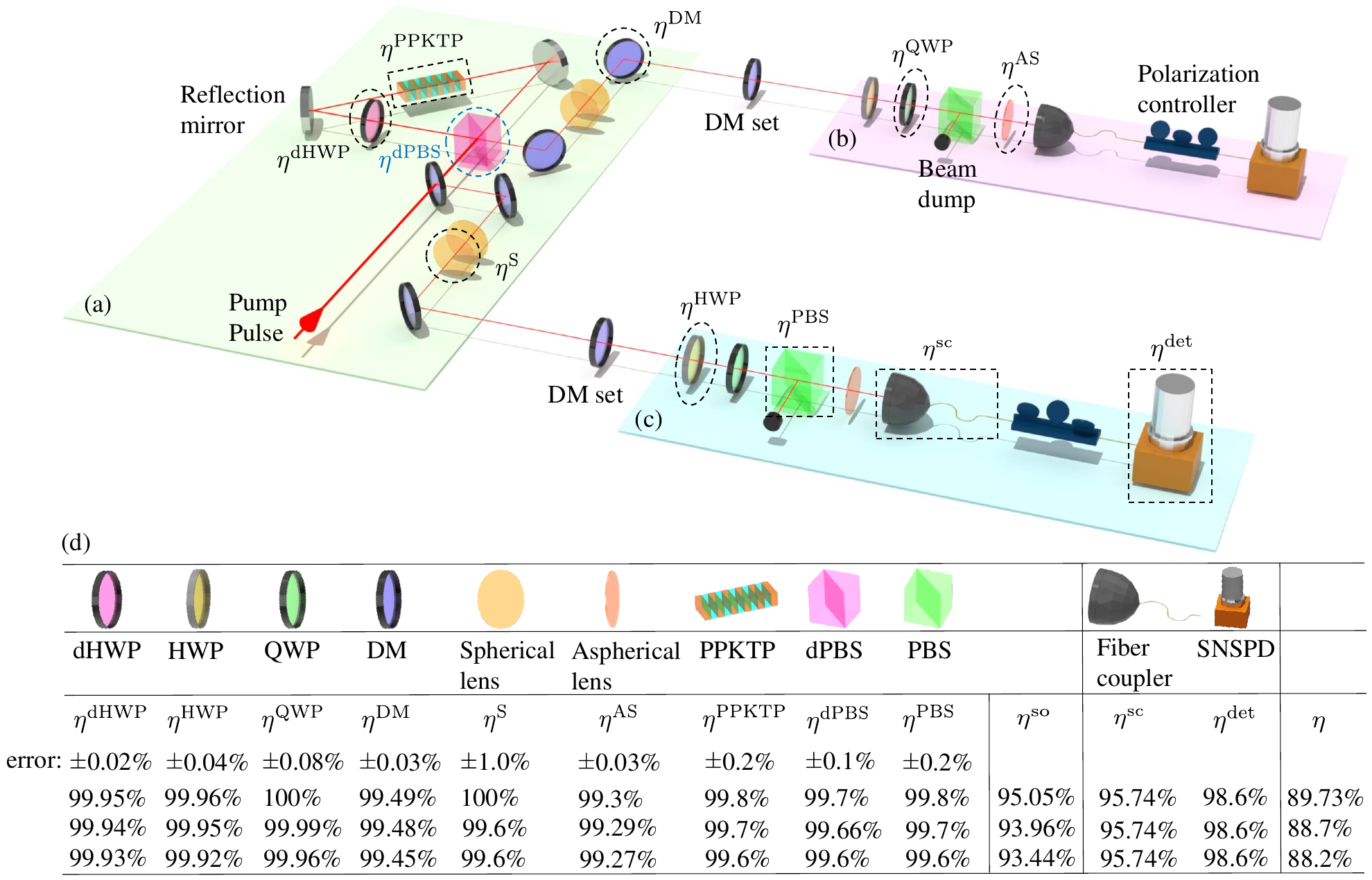}%jpg
\caption{
Schematic of the expected photonic experimental setup for our protocol.
(a) Alice's generation of entangled photon pairs.
(b) Alice's single-photon polarization projection measurement.
(c) Bob's single-photon polarization projection measurement.
(d) The table shows the detection efficiency $\eta$ which includes the efficiency $\eta^\text{sc}$ of single mode optical fiber coupler, the efficiency of SNSPD $\eta^\text{det}$, and the optical efficiency $\eta^\text{so}$ of dHWP, HWP, QWP, DM, spherical lens, aspherical lens, PPKTP, dPBS, and PBS.
Considering the errors of efficiencies of optical elements shown in the first row, the best detection efficiency $\eta$, which can be achieved with experimental setup in Ref.~\cite{Liu21} is $89.73\%$, as shown in the second row.
The efficiency values in the third (fourth) row are considered for the required detection efficiency $88.7\%$ ($88.2\%$) in our protocol.
}\label{experimentsetup}
\end{figure*}

The generation of entangled photon pairs is shown in Fig.~\ref{experimentsetup}(a). Alice injects light pulses into a periodically poled potassium titanyl phosphate (PPKTP) crystal in a Sagnac loop to generate polarization-entangled photon pairs.
Alice's single-photon polarization projection measurement is shown in Fig.~\ref{experimentsetup}(b). Alice projects the single photon into predetermined measurement bases using a set of half-wave plate (HWP) and quarter-wave plate (QWP). Then, the single photon is collected, transmitted through a fiber, and detected by a superconducting nanowire single-photon detector (SNSPD).
Bob's single-photon polarization projection measurement is shown in Fig.~\ref{experimentsetup}(c).
The detection efficiency $\eta$ derives from the efficiency of coupling entangled photons into single mode optical fiber $\eta^\text{sc}$,
the efficiency of the SNSPD $\eta^\text{det}$, and the optical efficiency due to limited transmittance of optical elements $\eta^\text{so}$,
i.e., $\eta = \eta^\text{sc} \eta^\text{det} \eta^\text{so}$ (where symbols of efficiencies here and below are from Ref.~\cite{Liu21}).
Furthermore, $\eta^\text{so} = \eta^\text{dHWP} \eta^\text{HWP} \eta^\text{QWP} (\eta^\text{DM})^7 (\eta^\text{S})^2 \eta^\text{AS} \eta^\text{PPKTP} \eta^\text{PBS} \eta^\text{dPBS}$, where
$\eta^\text{dHWP}$, $\eta^\text{HWP}$, $\eta^\text{QWP}$, $ \eta^\text{DM}$, $\eta^\text{S}$, $\eta^\text{AS}$, $\eta^\text{PPKTP}$, $\eta^\text{PBS}$, $\eta^\text{dPBS}$ denote the efficiencies of dual-wavelength HWP (dHWP), HWP, QWP, dichoric mirror (DM), spherical lens, aspherical lens, PPKTP, polarizing beam splitter (PBS), and dual-wavelength PBS (dPBS), respectively [as shown in Fig.~\ref{experimentsetup}(d)].
We assumed that the detection efficiency for Alice is equal to that for Bob, i.e., $\eta_A = \eta_B = \eta$, $\eta^\text{sc}_A = \eta^\text{sc}_B = \eta^\text{sc}$, $\eta^\text{det}_A = \eta^\text{det}_B = \eta^\text{det}$, and $\eta^\text{so}_A = \eta^\text{so}_B = \eta^\text{so}$.
Additionally, we supposed that the DM set for Alice is five DMs and for Bob is four DMs in Fig.~\ref{experimentsetup}. At the same time, the number of DMs is dependent on the experimental requirement in practice.
As shown in Fig.~\ref{experimentsetup}(d), the efficiency values in the third (fourth) row are considered for the required detection efficiency $88.7\%$ ($88.2\%$) in our protocol.

\section{COMPARISON OF KEY RATES WITH DIFFERENT DETECTION EFFICIENCIES AND NUMBERS OF KEY ROUNDS}\label{appendix:6}

\begin{figure}[t]
\includegraphics[width=8cm]{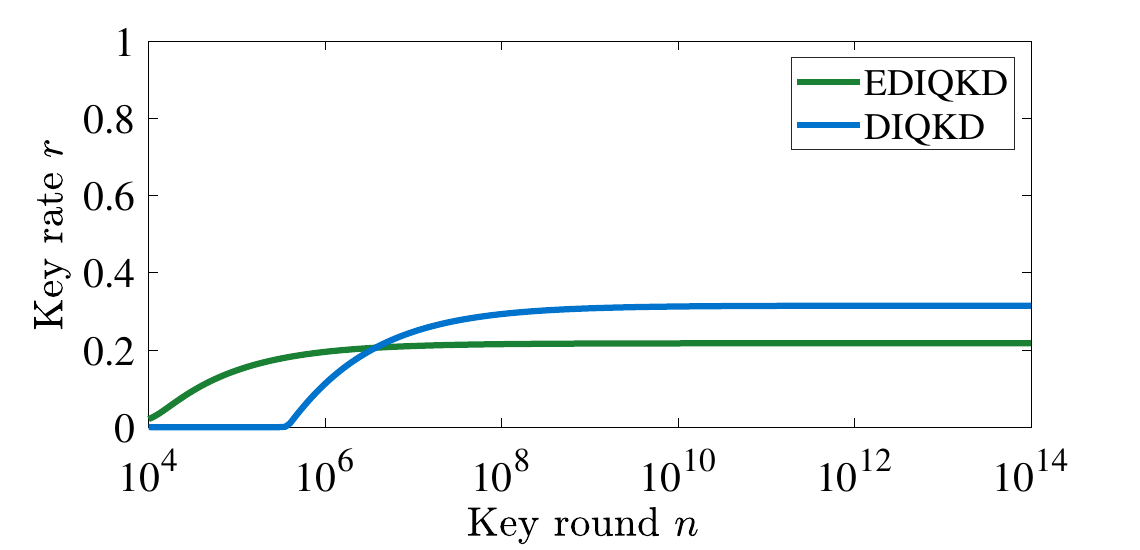}
\caption{
Comparison between key rates in EDIQKD and DIQKD protocols with varying number of key rounds under the detection efficiency $92\%$.
The key rate in EDIQKD (DIQKD) protocol is plotted by the dark green (blue) line, and the efficiency of EDIQKD protocol is $10^{2.04} \sim 109.65$ times higher than that of DIQKD protocol.
}\label{3D_b}
\end{figure}

\begin{table}[t]
\caption{
In this table, we show some examples (for the detection efficiency $88.8\% \leq \eta \leq 100\%$) of efficiency factors $E_\text{f}$,
which is obtained from the minimum number of key round to guarantee security in DIQKD protocol, $n_{\text{DIQKD}}'$, divided by which in EDIQKD protocol, $n_{\text{EDIQKD}}'$ (key rate $r \geq 0.001$ is chosen as the nonzero key rate in the numerical analysis).
}
\begin{ruledtabular}
\begin{tabular}{llll}
$\eta$ & $n_{\text{EDIQKD}}'$ & $n_{\text{DIQKD}}'$ & $E_\text{f} = n_{\text{DIQKD}}' / n_{\text{EDIQKD}}'$\\
[0.2em]
\hline\\
[-0.9em]
$100\%$ & $10^{2.21}$ & $10^{4.77}$ & $10^{2.56} \sim 363.01$\\
$95\%$ & $10^{2.88}$ & $10^{5.18}$ & $10^{2.30} \sim 199.53$\\
$92\%$ & $10^{3.48}$ & $10^{5.52}$ & $10^{2.04} \sim 109.65$\\
%
%$91\%$ & $10^{3.81}$ & $10^{5.66}$ & $10^{1.85} \sim 70.79$\\
$90\%$ & $10^{4.32}$ & $10^{5.82}$ & $10^{1.50} \sim 31.62$\\
$89.73\%$ & $10^{4.42}$ & $10^{5.86}$ & $10^{1.44} \sim 27.54$\\
%
%$89\%$ & $10^{5.42}$ & $10^{6.01}$ & $10^{0.59} \sim 3.89$\\
$88.9\%$ & $10^{5.48}$ & $10^{6.03}$ & $10^{0.55} \sim 3.55$\\
$88.8\%$ & $10^{6.08}$ & $10^{6.05}$ & $10^{-0.03} \sim 0.933$\\
\end{tabular}
\end{ruledtabular}
\label{3D_b_table}
\end{table}

To show how more efficient EDIQKD protocol than DIQKD protocol~\cite{Murta19} is to guarantee security for the imperfect detection efficiency, we execute the simulation similar to that in Fig.~\ref{keyratedetecteff}, but key rates are not maximized in the optimization problems with varying experimental parameters.
In addition to the critical factors in experimental imperfections considered in Ref.~\cite{Liu21}, we assumed that the ideally prepared entangled state is $\ket{\psi} = (\ket{0}\otimes\ket{1}+\ket{1}\otimes\ket{0})/\sqrt{2}$ with the best-entangled state fidelity $99.8\%$~\cite{Liu21}, the number of spontaneous parametric down-conversion entangled photon pair is $0.01$, and the chosen measurement outcomes are conditioned on the detected events of Alice's detector.
Considering that the chosen measurement settings satisfy the correlation between the QBER $Q$ and the value of CHSH polynomial $S$~\cite{Acin07}, i.e., $S = 2\sqrt{2}(1-2Q)$,
we choose $\{\hat{V}_{2}'^{(A)}, \hat{V}_{3}'^{(B)}\}$ as the set of measurements in key rounds and $\{\hat{V}_{1}'^{(A)}, \hat{V}_{2}'^{(A)}, \hat{V}_{1}'^{(B)}, \hat{V}_{2}'^{(B)}\}$ as the set of measurements in test rounds for DIQKD protocol \cite{Murta19},
where $\hat{V}_{2}'^{(A)} = \hat{V}_{3}'^{(B)} = Z$, $\hat{V}_{1}'^{(A)} = X$, $\hat{V}_{1}'^{(B)} = (X-Z)/\sqrt{2}$, $\hat{V}_{2}'^{(B)} = (X+Z)/\sqrt{2}$, and $X$ and $Z$ are the Pauli matrices.

The comparison between key rates in two protocols with numbers of key rounds under different detection efficiencies is shown in Fig.~\ref{3D}, and we take some detection efficiencies $\eta$ as examples for efficiency factors $E_\text{f}$ shown in Table~\ref{3D_b_table}.
The required detection efficiency in EDIQKD protocol is a little higher than that of DIQKD protocol with the number of key rounds greater than $10^{6}$,
However, the key rate in the EDIQKD protocol rises faster with the increase of detection efficiency and finally surpasses the key rate in the DIQKD protocol.
With the required detection efficiency $88.9\%$ and the number of key rounds less than $10^{6}$, the key rate in EDIQKD protocol is higher than that in DIQKD protocol, and the difference becomes more evident for the greater detection efficiency [see Fig.~\ref{3D}(a)].
As the detection efficiency is $92\%$, the minimum number of key rounds required to guarantee security (key rate $r \geq 0.001$ is chosen as the nonzero key rate in the numerical analysis) in DIQKD protocol is $10^{2.04} \sim 109.65$ times more than that in our protocol, so the efficiency of EDIQKD protocol is about 100 times higher than that of DIQKD protocol, as shown in Fig.~\ref{3D_b}.
It is worth noting that considering the best detection efficiency $89.73\%$ which can be achieved with experimental setup in Ref.~\cite{Liu21} [see Fig.~\ref{experimentsetup}(d) for details], the efficiency of EDIQKD protocol is $10^{1.44} \sim 27.54$ times higher than that of DIQKD protocol, as shown in Table~\ref{3D_b_table}.

%%%  Bibliography %%%

\bibliography{DIPaMQKD}

\begin{thebibliography}{36}%
\makeatletter
\providecommand \@ifxundefined [1]{%
 \@ifx{#1\undefined}
}%
\providecommand \@ifnum [1]{%
 \ifnum #1\expandafter \@firstoftwo
 \else \expandafter \@secondoftwo
 \fi
}%
\providecommand \@ifx [1]{%
 \ifx #1\expandafter \@firstoftwo
 \else \expandafter \@secondoftwo
 \fi
}%
\providecommand \natexlab [1]{#1}%
\providecommand \enquote  [1]{``#1''}%
\providecommand \bibnamefont  [1]{#1}%
\providecommand \bibfnamefont [1]{#1}%
\providecommand \citenamefont [1]{#1}%
\providecommand \href@noop [0]{\@secondoftwo}%
\providecommand \href [0]{\begingroup \@sanitize@url \@href}%
\providecommand \@href[1]{\@@startlink{#1}\@@href}%
\providecommand \@@href[1]{\endgroup#1\@@endlink}%
\providecommand \@sanitize@url [0]{\catcode `\\12\catcode `\$12\catcode
  `\&12\catcode `\#12\catcode `\^12\catcode `\_12\catcode `\%12\relax}%
\providecommand \@@startlink[1]{}%
\providecommand \@@endlink[0]{}%
\providecommand \url  [0]{\begingroup\@sanitize@url \@url }%
\providecommand \@url [1]{\endgroup\@href {#1}{\urlprefix }}%
\providecommand \urlprefix  [0]{URL }%
\providecommand \Eprint [0]{\href }%
\providecommand \doibase [0]{https://doi.org/}%
\providecommand \selectlanguage [0]{\@gobble}%
\providecommand \bibinfo  [0]{\@secondoftwo}%
\providecommand \bibfield  [0]{\@secondoftwo}%
\providecommand \translation [1]{[#1]}%
\providecommand \BibitemOpen [0]{}%
\providecommand \bibitemStop [0]{}%
\providecommand \bibitemNoStop [0]{.\EOS\space}%
\providecommand \EOS [0]{\spacefactor3000\relax}%
\providecommand \BibitemShut  [1]{\csname bibitem#1\endcsname}%
\let\auto@bib@innerbib\@empty
%</preamble>
\bibitem [{\citenamefont {Gisin}\ and\ \citenamefont
  {Thew}(2007)}]{gisin2007quantum}%
  \BibitemOpen
  \bibfield  {author} {\bibinfo {author} {\bibfnamefont {N.}~\bibnamefont
  {Gisin}}\ and\ \bibinfo {author} {\bibfnamefont {R.}~\bibnamefont {Thew}},\
  }\bibfield  {title} {\bibinfo {title} {Quantum communication},\ }\href@noop
  {} {\bibfield  {journal} {\bibinfo  {journal} {Nature photonics}\ }\textbf
  {\bibinfo {volume} {1}},\ \bibinfo {pages} {165} (\bibinfo {year}
  {2007})}\BibitemShut {NoStop}%
\bibitem [{\citenamefont {Lo}\ \emph {et~al.}(2014)\citenamefont {Lo},
  \citenamefont {Curty},\ and\ \citenamefont {Tamaki}}]{lo2014secure}%
  \BibitemOpen
  \bibfield  {author} {\bibinfo {author} {\bibfnamefont {H.-K.}\ \bibnamefont
  {Lo}}, \bibinfo {author} {\bibfnamefont {M.}~\bibnamefont {Curty}},\ and\
  \bibinfo {author} {\bibfnamefont {K.}~\bibnamefont {Tamaki}},\ }\bibfield
  {title} {\bibinfo {title} {Secure quantum key distribution},\ }\href@noop {}
  {\bibfield  {journal} {\bibinfo  {journal} {Nature Photonics}\ }\textbf
  {\bibinfo {volume} {8}},\ \bibinfo {pages} {595} (\bibinfo {year}
  {2014})}\BibitemShut {NoStop}%
\bibitem [{\citenamefont {Diamanti}\ \emph {et~al.}(2016)\citenamefont
  {Diamanti}, \citenamefont {Lo}, \citenamefont {Qi},\ and\ \citenamefont
  {Yuan}}]{diamanti2016practical}%
  \BibitemOpen
  \bibfield  {author} {\bibinfo {author} {\bibfnamefont {E.}~\bibnamefont
  {Diamanti}}, \bibinfo {author} {\bibfnamefont {H.-K.}\ \bibnamefont {Lo}},
  \bibinfo {author} {\bibfnamefont {B.}~\bibnamefont {Qi}},\ and\ \bibinfo
  {author} {\bibfnamefont {Z.}~\bibnamefont {Yuan}},\ }\bibfield  {title}
  {\bibinfo {title} {Practical challenges in quantum key distribution},\
  }\href@noop {} {\bibfield  {journal} {\bibinfo  {journal} {npj Quantum
  Information}\ }\textbf {\bibinfo {volume} {2}},\ \bibinfo {pages} {1}
  (\bibinfo {year} {2016})}\BibitemShut {NoStop}%
\bibitem [{\citenamefont {Pirandola}\ \emph {et~al.}(2020)\citenamefont
  {Pirandola}, \citenamefont {Andersen}, \citenamefont {Banchi}, \citenamefont
  {Berta}, \citenamefont {Bunandar}, \citenamefont {Colbeck}, \citenamefont
  {Englund}, \citenamefont {Gehring}, \citenamefont {Lupo}, \citenamefont
  {Ottaviani} \emph {et~al.}}]{pirandola2020advances}%
  \BibitemOpen
  \bibfield  {author} {\bibinfo {author} {\bibfnamefont {S.}~\bibnamefont
  {Pirandola}}, \bibinfo {author} {\bibfnamefont {U.~L.}\ \bibnamefont
  {Andersen}}, \bibinfo {author} {\bibfnamefont {L.}~\bibnamefont {Banchi}},
  \bibinfo {author} {\bibfnamefont {M.}~\bibnamefont {Berta}}, \bibinfo
  {author} {\bibfnamefont {D.}~\bibnamefont {Bunandar}}, \bibinfo {author}
  {\bibfnamefont {R.}~\bibnamefont {Colbeck}}, \bibinfo {author} {\bibfnamefont
  {D.}~\bibnamefont {Englund}}, \bibinfo {author} {\bibfnamefont
  {T.}~\bibnamefont {Gehring}}, \bibinfo {author} {\bibfnamefont
  {C.}~\bibnamefont {Lupo}}, \bibinfo {author} {\bibfnamefont {C.}~\bibnamefont
  {Ottaviani}}, \emph {et~al.},\ }\bibfield  {title} {\bibinfo {title}
  {Advances in quantum cryptography},\ }\href@noop {} {\bibfield  {journal}
  {\bibinfo  {journal} {Advances in optics and photonics}\ }\textbf {\bibinfo
  {volume} {12}},\ \bibinfo {pages} {1012} (\bibinfo {year}
  {2020})}\BibitemShut {NoStop}%
\bibitem [{\citenamefont {Ekert}(1991)}]{Ekert91}%
  \BibitemOpen
  \bibfield  {author} {\bibinfo {author} {\bibfnamefont {A.~K.}\ \bibnamefont
  {Ekert}},\ }\bibfield  {title} {\bibinfo {title} {Quantum cryptography based
  on {B}ell's theorem},\ }\href {https://doi.org/10.1103/PhysRevLett.67.661}
  {\bibfield  {journal} {\bibinfo  {journal} {Phys. Rev. Lett.}\ }\textbf
  {\bibinfo {volume} {67}},\ \bibinfo {pages} {661} (\bibinfo {year}
  {1991})}\BibitemShut {NoStop}%
\bibitem [{\citenamefont {Bennett}\ and\ \citenamefont
  {Brassard}(1984)}]{Bennett84}%
  \BibitemOpen
  \bibfield  {author} {\bibinfo {author} {\bibfnamefont {C.~H.}\ \bibnamefont
  {Bennett}}\ and\ \bibinfo {author} {\bibfnamefont {G.}~\bibnamefont
  {Brassard}},\ }\bibfield  {title} {\bibinfo {title} {Quantum cryptography:
  Public key distribution and coin tossing},\ }in\ \href@noop {} {\emph
  {\bibinfo {booktitle} {Proceedings of IEEE International Conference on
  Computers, Systems and Signal Processing}}}\ (\bibinfo {year} {1984})\ pp.\
  \bibinfo {pages} {175--179}\BibitemShut {NoStop}%
\bibitem [{\citenamefont {Pironio}\ \emph {et~al.}(2009)\citenamefont
  {Pironio}, \citenamefont {Acín}, \citenamefont {Brunner}, \citenamefont
  {Gisin}, \citenamefont {Massar},\ and\ \citenamefont {Scarani}}]{Pironio09}%
  \BibitemOpen
  \bibfield  {author} {\bibinfo {author} {\bibfnamefont {S.}~\bibnamefont
  {Pironio}}, \bibinfo {author} {\bibfnamefont {A.}~\bibnamefont {Acín}},
  \bibinfo {author} {\bibfnamefont {N.}~\bibnamefont {Brunner}}, \bibinfo
  {author} {\bibfnamefont {N.}~\bibnamefont {Gisin}}, \bibinfo {author}
  {\bibfnamefont {S.}~\bibnamefont {Massar}},\ and\ \bibinfo {author}
  {\bibfnamefont {V.}~\bibnamefont {Scarani}},\ }\bibfield  {title} {\bibinfo
  {title} {Device-independent quantum key distribution secure against
  collective attacks},\ }\href {https://doi.org/10.1088/1367-2630/11/4/045021}
  {\bibfield  {journal} {\bibinfo  {journal} {New. J. Phys.}\ }\textbf
  {\bibinfo {volume} {11}},\ \bibinfo {pages} {045021} (\bibinfo {year}
  {2009})}\BibitemShut {NoStop}%
\bibitem [{\citenamefont {Xu}\ \emph {et~al.}(2020)\citenamefont {Xu},
  \citenamefont {Ma}, \citenamefont {Zhang}, \citenamefont {Lo},\ and\
  \citenamefont {Pan}}]{Xu20}%
  \BibitemOpen
  \bibfield  {author} {\bibinfo {author} {\bibfnamefont {F.}~\bibnamefont
  {Xu}}, \bibinfo {author} {\bibfnamefont {X.}~\bibnamefont {Ma}}, \bibinfo
  {author} {\bibfnamefont {Q.}~\bibnamefont {Zhang}}, \bibinfo {author}
  {\bibfnamefont {H.-K.}\ \bibnamefont {Lo}},\ and\ \bibinfo {author}
  {\bibfnamefont {J.-W.}\ \bibnamefont {Pan}},\ }\bibfield  {title} {\bibinfo
  {title} {Secure quantum key distribution with realistic devices},\ }\href
  {https://doi.org/10.1103/RevModPhys.92.025002} {\bibfield  {journal}
  {\bibinfo  {journal} {Rev. Mod. Phys.}\ }\textbf {\bibinfo {volume} {92}},\
  \bibinfo {pages} {025002} (\bibinfo {year} {2020})}\BibitemShut {NoStop}%
\bibitem [{\citenamefont {Ac\'{\i}n}\ \emph {et~al.}(2007)\citenamefont
  {Ac\'{\i}n}, \citenamefont {Brunner}, \citenamefont {Gisin}, \citenamefont
  {Massar}, \citenamefont {Pironio},\ and\ \citenamefont {Scarani}}]{Acin07}%
  \BibitemOpen
  \bibfield  {author} {\bibinfo {author} {\bibfnamefont {A.}~\bibnamefont
  {Ac\'{\i}n}}, \bibinfo {author} {\bibfnamefont {N.}~\bibnamefont {Brunner}},
  \bibinfo {author} {\bibfnamefont {N.}~\bibnamefont {Gisin}}, \bibinfo
  {author} {\bibfnamefont {S.}~\bibnamefont {Massar}}, \bibinfo {author}
  {\bibfnamefont {S.}~\bibnamefont {Pironio}},\ and\ \bibinfo {author}
  {\bibfnamefont {V.}~\bibnamefont {Scarani}},\ }\bibfield  {title} {\bibinfo
  {title} {Device-independent security of quantum cryptography against
  collective attacks},\ }\href {https://doi.org/10.1103/PhysRevLett.98.230501}
  {\bibfield  {journal} {\bibinfo  {journal} {Phys. Rev. Lett.}\ }\textbf
  {\bibinfo {volume} {98}},\ \bibinfo {pages} {230501} (\bibinfo {year}
  {2007})}\BibitemShut {NoStop}%
\bibitem [{\citenamefont {Primaatmaja}\ \emph {et~al.}(2023)\citenamefont
  {Primaatmaja}, \citenamefont {Goh}, \citenamefont {Tan}, \citenamefont
  {Khoo}, \citenamefont {Ghorai},\ and\ \citenamefont
  {Lim}}]{Primaatmaja2023securityofdevice}%
  \BibitemOpen
  \bibfield  {author} {\bibinfo {author} {\bibfnamefont {I.~W.}\ \bibnamefont
  {Primaatmaja}}, \bibinfo {author} {\bibfnamefont {K.~T.}\ \bibnamefont
  {Goh}}, \bibinfo {author} {\bibfnamefont {E.~Y.-Z.}\ \bibnamefont {Tan}},
  \bibinfo {author} {\bibfnamefont {J.~T.-F.}\ \bibnamefont {Khoo}}, \bibinfo
  {author} {\bibfnamefont {S.}~\bibnamefont {Ghorai}},\ and\ \bibinfo {author}
  {\bibfnamefont {C.~C.-W.}\ \bibnamefont {Lim}},\ }\bibfield  {title}
  {\bibinfo {title} {Security of device-independent quantum key distribution
  protocols: a review},\ }\href {https://doi.org/10.22331/q-2023-03-02-932}
  {\bibfield  {journal} {\bibinfo  {journal} {{Quantum}}\ }\textbf {\bibinfo
  {volume} {7}},\ \bibinfo {pages} {932} (\bibinfo {year} {2023})}\BibitemShut
  {NoStop}%
\bibitem [{\citenamefont {Zapatero}\ \emph {et~al.}(2023)\citenamefont
  {Zapatero}, \citenamefont {van Leent}, \citenamefont {Arnon-Friedman},
  \citenamefont {Liu}, \citenamefont {Zhang}, \citenamefont {Weinfurter},\ and\
  \citenamefont {Curty}}]{Zapatero2023}%
  \BibitemOpen
  \bibfield  {author} {\bibinfo {author} {\bibfnamefont {V.}~\bibnamefont
  {Zapatero}}, \bibinfo {author} {\bibfnamefont {T.}~\bibnamefont {van Leent}},
  \bibinfo {author} {\bibfnamefont {R.}~\bibnamefont {Arnon-Friedman}},
  \bibinfo {author} {\bibfnamefont {W.-Z.}\ \bibnamefont {Liu}}, \bibinfo
  {author} {\bibfnamefont {Q.}~\bibnamefont {Zhang}}, \bibinfo {author}
  {\bibfnamefont {H.}~\bibnamefont {Weinfurter}},\ and\ \bibinfo {author}
  {\bibfnamefont {M.}~\bibnamefont {Curty}},\ }\bibfield  {title} {\bibinfo
  {title} {Advances in device-independent quantum key distribution},\
  }\href@noop {} {\bibfield  {journal} {\bibinfo  {journal} {npj Quantum Inf}\
  }\textbf {\bibinfo {volume} {9}},\ \bibinfo {pages} {10} (\bibinfo {year}
  {2023})}\BibitemShut {NoStop}%
\bibitem [{\citenamefont {Chen}\ \emph {et~al.}(2020)\citenamefont {Chen},
  \citenamefont {Lu}, \citenamefont {Sun}, \citenamefont {Zhang}, \citenamefont
  {Chen},\ and\ \citenamefont {Li}}]{Chen20}%
  \BibitemOpen
  \bibfield  {author} {\bibinfo {author} {\bibfnamefont {S.-H.}\ \bibnamefont
  {Chen}}, \bibinfo {author} {\bibfnamefont {H.}~\bibnamefont {Lu}}, \bibinfo
  {author} {\bibfnamefont {Q.-C.}\ \bibnamefont {Sun}}, \bibinfo {author}
  {\bibfnamefont {Q.}~\bibnamefont {Zhang}}, \bibinfo {author} {\bibfnamefont
  {Y.-A.}\ \bibnamefont {Chen}},\ and\ \bibinfo {author} {\bibfnamefont
  {C.-M.}\ \bibnamefont {Li}},\ }\bibfield  {title} {\bibinfo {title}
  {Discriminating quantum correlations with networking quantum teleportation},\
  }\href {https://doi.org/10.1103/PhysRevResearch.2.013043} {\bibfield
  {journal} {\bibinfo  {journal} {Phys. Rev. Res.}\ }\textbf {\bibinfo {volume}
  {2}},\ \bibinfo {pages} {013043} (\bibinfo {year} {2020})}\BibitemShut
  {NoStop}%
\bibitem [{\citenamefont {Chuang}\ and\ \citenamefont
  {Nielsen}(1997)}]{Chuang96}%
  \BibitemOpen
  \bibfield  {author} {\bibinfo {author} {\bibfnamefont {I.~L.}\ \bibnamefont
  {Chuang}}\ and\ \bibinfo {author} {\bibfnamefont {M.~A.}\ \bibnamefont
  {Nielsen}},\ }\bibfield  {title} {\bibinfo {title} {Prescription for
  experimental determination of the dynamics of a quantum black box},\ }\href
  {https://doi.org/10.1080/09500349708231894} {\bibfield  {journal} {\bibinfo
  {journal} {J. Mod. Opt.}\ }\textbf {\bibinfo {volume} {44}},\ \bibinfo
  {pages} {2455} (\bibinfo {year} {1997})}\BibitemShut {NoStop}%
\bibitem [{\citenamefont {Nielsen}\ and\ \citenamefont
  {Chuang}(2000)}]{Nielsen&Chuang00}%
  \BibitemOpen
  \bibfield  {author} {\bibinfo {author} {\bibfnamefont {M.~A.}\ \bibnamefont
  {Nielsen}}\ and\ \bibinfo {author} {\bibfnamefont {I.~L.}\ \bibnamefont
  {Chuang}},\ }\href@noop {} {\emph {\bibinfo {title} {Quantum Computation and
  Quantum Information}}}\ (\bibinfo  {publisher} {Cambridge Univ. Press},\
  \bibinfo {year} {2000})\BibitemShut {NoStop}%
\bibitem [{\citenamefont {Lofberg}(2004)}]{Lofberg}%
  \BibitemOpen
  \bibfield  {author} {\bibinfo {author} {\bibfnamefont {J.}~\bibnamefont
  {Lofberg}},\ }\bibfield  {title} {\bibinfo {title} {{YALMIP}: {A} toolbox for
  modeling and optimization in {MATLAB}},\ }in\ \href
  {https://doi.org/10.1109/CACSD.2004.1393890} {\emph {\bibinfo {booktitle}
  {2004 IEEE International Conference on Robotics and Automation (IEEE Cat.
  No.04CH37508)}}}\ (\bibinfo {year} {2004})\ pp.\ \bibinfo {pages}
  {284--289}\BibitemShut {NoStop}%
\bibitem [{\citenamefont {Toh}\ \emph {et~al.}(1999)\citenamefont {Toh},
  \citenamefont {Todd},\ and\ \citenamefont {T\"{u}t\"{u}nc\"{u}}}]{sdpsolver}%
  \BibitemOpen
  \bibfield  {author} {\bibinfo {author} {\bibfnamefont {K.~C.}\ \bibnamefont
  {Toh}}, \bibinfo {author} {\bibfnamefont {M.~J.}\ \bibnamefont {Todd}},\ and\
  \bibinfo {author} {\bibfnamefont {R.~H.}\ \bibnamefont
  {T\"{u}t\"{u}nc\"{u}}},\ }\bibfield  {title} {\bibinfo {title} {{SDPT3} --
  {A} {MATLAB} software package for semidefinite programming},\ }\href@noop {}
  {\bibfield  {journal} {\bibinfo  {journal} {Optim. Methods Softw.}\ }\textbf
  {\bibinfo {volume} {11}},\ \bibinfo {pages} {545} (\bibinfo {year}
  {1999})}\BibitemShut {NoStop}%
\bibitem [{\citenamefont {Murta}\ \emph {et~al.}(2019)\citenamefont {Murta},
  \citenamefont {van Dam}, \citenamefont {Ribeiro}, \citenamefont {Hanson},\
  and\ \citenamefont {Wehner}}]{Murta19}%
  \BibitemOpen
  \bibfield  {author} {\bibinfo {author} {\bibfnamefont {G.}~\bibnamefont
  {Murta}}, \bibinfo {author} {\bibfnamefont {S.~B.}\ \bibnamefont {van Dam}},
  \bibinfo {author} {\bibfnamefont {J.}~\bibnamefont {Ribeiro}}, \bibinfo
  {author} {\bibfnamefont {R.}~\bibnamefont {Hanson}},\ and\ \bibinfo {author}
  {\bibfnamefont {S.}~\bibnamefont {Wehner}},\ }\bibfield  {title} {\bibinfo
  {title} {Towards a realization of device-independent quantum key
  distribution},\ }\href {https://doi.org/10.1088/2058-9565/ab2819} {\bibfield
  {journal} {\bibinfo  {journal} {Quantum Sci. Technol.}\ }\textbf {\bibinfo
  {volume} {4}},\ \bibinfo {pages} {035011} (\bibinfo {year}
  {2019})}\BibitemShut {NoStop}%
\bibitem [{\citenamefont {Iqbal}\ \emph {et~al.}(2022)\citenamefont {Iqbal},
  \citenamefont {Velasco}, \citenamefont {Ruiz}, \citenamefont {Napoli},
  \citenamefont {Pedro},\ and\ \citenamefont {Costa}}]{Masab22}%
  \BibitemOpen
  \bibfield  {author} {\bibinfo {author} {\bibfnamefont {M.}~\bibnamefont
  {Iqbal}}, \bibinfo {author} {\bibfnamefont {L.}~\bibnamefont {Velasco}},
  \bibinfo {author} {\bibfnamefont {M.}~\bibnamefont {Ruiz}}, \bibinfo {author}
  {\bibfnamefont {A.}~\bibnamefont {Napoli}}, \bibinfo {author} {\bibfnamefont
  {J.}~\bibnamefont {Pedro}},\ and\ \bibinfo {author} {\bibfnamefont
  {N.}~\bibnamefont {Costa}},\ }\bibfield  {title} {\bibinfo {title} {Quantum
  bit retransmission using universal quantum copying machine},\ }in\ \href
  {https://doi.org/10.23919/ONDM54585.2022.9782866} {\emph {\bibinfo
  {booktitle} {2022 International Conference on Optical Network Design and
  Modeling (ONDM)}}}\ (\bibinfo {year} {2022})\ pp.\ \bibinfo {pages}
  {1--3}\BibitemShut {NoStop}%
\bibitem [{\citenamefont {Scarani}\ \emph {et~al.}(2005)\citenamefont
  {Scarani}, \citenamefont {Iblisdir}, \citenamefont {Gisin},\ and\
  \citenamefont {Ac\'{\i}n}}]{Scarani05}%
  \BibitemOpen
  \bibfield  {author} {\bibinfo {author} {\bibfnamefont {V.}~\bibnamefont
  {Scarani}}, \bibinfo {author} {\bibfnamefont {S.}~\bibnamefont {Iblisdir}},
  \bibinfo {author} {\bibfnamefont {N.}~\bibnamefont {Gisin}},\ and\ \bibinfo
  {author} {\bibfnamefont {A.}~\bibnamefont {Ac\'{\i}n}},\ }\bibfield  {title}
  {\bibinfo {title} {Quantum cloning},\ }\href
  {https://doi.org/10.1103/RevModPhys.77.1225} {\bibfield  {journal} {\bibinfo
  {journal} {Rev. Mod. Phys.}\ }\textbf {\bibinfo {volume} {77}},\ \bibinfo
  {pages} {1225} (\bibinfo {year} {2005})}\BibitemShut {NoStop}%
\bibitem [{\citenamefont {Ferenczi}\ and\ \citenamefont
  {L\"utkenhaus}(2012)}]{PhysRevA.85.052310}%
  \BibitemOpen
  \bibfield  {author} {\bibinfo {author} {\bibfnamefont {A.}~\bibnamefont
  {Ferenczi}}\ and\ \bibinfo {author} {\bibfnamefont {N.}~\bibnamefont
  {L\"utkenhaus}},\ }\bibfield  {title} {\bibinfo {title} {Symmetries in
  quantum key distribution and the connection between optimal attacks and
  optimal cloning},\ }\href@noop {} {\bibfield  {journal} {\bibinfo  {journal}
  {Phys. Rev. A}\ }\textbf {\bibinfo {volume} {85}},\ \bibinfo {pages} {052310}
  (\bibinfo {year} {2012})}\BibitemShut {NoStop}%
\bibitem [{\citenamefont {Chiu}\ \emph {et~al.}(2016)\citenamefont {Chiu},
  \citenamefont {Lambert}, \citenamefont {Liao}, \citenamefont {Nori},\ and\
  \citenamefont {Li}}]{Chiu16}%
  \BibitemOpen
  \bibfield  {author} {\bibinfo {author} {\bibfnamefont {C.-Y.}\ \bibnamefont
  {Chiu}}, \bibinfo {author} {\bibfnamefont {N.}~\bibnamefont {Lambert}},
  \bibinfo {author} {\bibfnamefont {T.-L.}\ \bibnamefont {Liao}}, \bibinfo
  {author} {\bibfnamefont {F.}~\bibnamefont {Nori}},\ and\ \bibinfo {author}
  {\bibfnamefont {C.-M.}\ \bibnamefont {Li}},\ }\bibfield  {title} {\bibinfo
  {title} {No-cloning of quantum steering},\ }\href@noop {} {\bibfield
  {journal} {\bibinfo  {journal} {npj Quantum Inf.}\ }\textbf {\bibinfo
  {volume} {2}},\ \bibinfo {pages} {16020} (\bibinfo {year}
  {2016})}\BibitemShut {NoStop}%
\bibitem [{\citenamefont {Devetak}\ and\ \citenamefont
  {A.Winter}(2005)}]{Devetak05}%
  \BibitemOpen
  \bibfield  {author} {\bibinfo {author} {\bibfnamefont {I.}~\bibnamefont
  {Devetak}}\ and\ \bibinfo {author} {\bibnamefont {A.Winter}},\ }\bibfield
  {title} {\bibinfo {title} {Distillation of secret key and entanglement from
  quantum states},\ }\href@noop {} {\bibfield  {journal} {\bibinfo  {journal}
  {Proc. R. Soc. A}\ }\textbf {\bibinfo {volume} {461}},\ \bibinfo {pages}
  {207} (\bibinfo {year} {2005})}\BibitemShut {NoStop}%
\bibitem [{\citenamefont {Sheridan}\ and\ \citenamefont
  {Scarani}(2010)}]{Sheridan10}%
  \BibitemOpen
  \bibfield  {author} {\bibinfo {author} {\bibfnamefont {L.}~\bibnamefont
  {Sheridan}}\ and\ \bibinfo {author} {\bibfnamefont {V.}~\bibnamefont
  {Scarani}},\ }\bibfield  {title} {\bibinfo {title} {Security proof for
  quantum key distribution using qudit systems},\ }\href
  {https://doi.org/10.1103/PhysRevA.82.030301} {\bibfield  {journal} {\bibinfo
  {journal} {Phys. Rev. A}\ }\textbf {\bibinfo {volume} {82}},\ \bibinfo
  {pages} {030301} (\bibinfo {year} {2010})}\BibitemShut {NoStop}%
\bibitem [{\citenamefont {Acín}\ \emph {et~al.}(2006)\citenamefont {Acín},
  \citenamefont {Massar},\ and\ \citenamefont {Pironio}}]{Acin06}%
  \BibitemOpen
  \bibfield  {author} {\bibinfo {author} {\bibfnamefont {A.}~\bibnamefont
  {Acín}}, \bibinfo {author} {\bibfnamefont {S.}~\bibnamefont {Massar}},\ and\
  \bibinfo {author} {\bibfnamefont {S.}~\bibnamefont {Pironio}},\ }\bibfield
  {title} {\bibinfo {title} {Efficient quantum key distribution secure against
  no-signalling eavesdroppers},\ }\href
  {https://doi.org/10.1088/1367-2630/8/8/126} {\bibfield  {journal} {\bibinfo
  {journal} {New J. Phys.}\ }\textbf {\bibinfo {volume} {8}},\ \bibinfo {pages}
  {126} (\bibinfo {year} {2006})}\BibitemShut {NoStop}%
\bibitem [{\citenamefont {Clauser}\ \emph {et~al.}(1969)\citenamefont
  {Clauser}, \citenamefont {Horne}, \citenamefont {Shimony},\ and\
  \citenamefont {Holt}}]{Clauser69}%
  \BibitemOpen
  \bibfield  {author} {\bibinfo {author} {\bibfnamefont {J.~F.}\ \bibnamefont
  {Clauser}}, \bibinfo {author} {\bibfnamefont {M.~A.}\ \bibnamefont {Horne}},
  \bibinfo {author} {\bibfnamefont {A.}~\bibnamefont {Shimony}},\ and\ \bibinfo
  {author} {\bibfnamefont {R.~A.}\ \bibnamefont {Holt}},\ }\bibfield  {title}
  {\bibinfo {title} {Proposed experiment to test local hidden-variable
  theories},\ }\href {https://doi.org/10.1103/PhysRevLett.23.880} {\bibfield
  {journal} {\bibinfo  {journal} {Phys. Rev. Lett.}\ }\textbf {\bibinfo
  {volume} {23}},\ \bibinfo {pages} {880} (\bibinfo {year} {1969})}\BibitemShut
  {NoStop}%
\bibitem [{\citenamefont {Arnon-Friedman}\ \emph {et~al.}(2018)\citenamefont
  {Arnon-Friedman}, \citenamefont {Dupuis}, \citenamefont {Fawzi},
  \citenamefont {Renner},\ and\ \citenamefont {Vidick}}]{Rotem2018}%
  \BibitemOpen
  \bibfield  {author} {\bibinfo {author} {\bibfnamefont {R.}~\bibnamefont
  {Arnon-Friedman}}, \bibinfo {author} {\bibfnamefont {F.}~\bibnamefont
  {Dupuis}}, \bibinfo {author} {\bibfnamefont {O.}~\bibnamefont {Fawzi}},
  \bibinfo {author} {\bibfnamefont {R.}~\bibnamefont {Renner}},\ and\ \bibinfo
  {author} {\bibfnamefont {T.}~\bibnamefont {Vidick}},\ }\bibfield  {title}
  {\bibinfo {title} {Practical device-independent quantum cryptography via
  entropy accumulation},\ }\href@noop {} {\bibfield  {journal} {\bibinfo
  {journal} {Nat. Commun.}\ }\textbf {\bibinfo {volume} {9}},\ \bibinfo {pages}
  {459} (\bibinfo {year} {2018})}\BibitemShut {NoStop}%
\bibitem [{\citenamefont {Tomamichel}\ and\ \citenamefont
  {Leverrier}(2017)}]{Tomamichel17}%
  \BibitemOpen
  \bibfield  {author} {\bibinfo {author} {\bibfnamefont {M.}~\bibnamefont
  {Tomamichel}}\ and\ \bibinfo {author} {\bibfnamefont {A.}~\bibnamefont
  {Leverrier}},\ }\bibfield  {title} {\bibinfo {title} {A largely
  self-contained and complete security proof for quantum key distribution},\
  }\href {https://doi.org/10.22331/q-2017-07-14-14} {\bibfield  {journal}
  {\bibinfo  {journal} {Quantum}\ }\textbf {\bibinfo {volume} {1}},\ \bibinfo
  {pages} {14} (\bibinfo {year} {2017})}\BibitemShut {NoStop}%
\bibitem [{\citenamefont {Liu}\ \emph {et~al.}(2022)\citenamefont {Liu},
  \citenamefont {Zhang}, \citenamefont {Zhen}, \citenamefont {Li},
  \citenamefont {Liu}, \citenamefont {Fan}, \citenamefont {Xu}, \citenamefont
  {Zhang},\ and\ \citenamefont {Pan}}]{Liu21}%
  \BibitemOpen
  \bibfield  {author} {\bibinfo {author} {\bibfnamefont {W.-Z.}\ \bibnamefont
  {Liu}}, \bibinfo {author} {\bibfnamefont {Y.-Z.}\ \bibnamefont {Zhang}},
  \bibinfo {author} {\bibfnamefont {Y.-Z.}\ \bibnamefont {Zhen}}, \bibinfo
  {author} {\bibfnamefont {M.-H.}\ \bibnamefont {Li}}, \bibinfo {author}
  {\bibfnamefont {Y.}~\bibnamefont {Liu}}, \bibinfo {author} {\bibfnamefont
  {J.}~\bibnamefont {Fan}}, \bibinfo {author} {\bibfnamefont {F.}~\bibnamefont
  {Xu}}, \bibinfo {author} {\bibfnamefont {Q.}~\bibnamefont {Zhang}},\ and\
  \bibinfo {author} {\bibfnamefont {J.-W.}\ \bibnamefont {Pan}},\ }\bibfield
  {title} {\bibinfo {title} {Toward a photonic demonstration of
  device-independent quantum key distribution},\ }\href
  {https://doi.org/10.1103/PhysRevLett.129.050502} {\bibfield  {journal}
  {\bibinfo  {journal} {Phys. Rev. Lett.}\ }\textbf {\bibinfo {volume} {129}},\
  \bibinfo {pages} {050502} (\bibinfo {year} {2022})}\BibitemShut {NoStop}%
\bibitem [{\citenamefont {Huang}\ \emph {et~al.}(2020)\citenamefont {Huang},
  \citenamefont {Xu},\ and\ \citenamefont {Li}}]{NiNi}%
  \BibitemOpen
  \bibfield  {author} {\bibinfo {author} {\bibfnamefont {N.-N.}\ \bibnamefont
  {Huang}}, \bibinfo {author} {\bibfnamefont {J.-C.}\ \bibnamefont {Xu}},\ and\
  \bibinfo {author} {\bibfnamefont {C.-M.}\ \bibnamefont {Li}},\ }\bibfield
  {title} {\bibinfo {title} {Device-independent quantum computing},\ }in\ \href
  {https://doi.org/10.1364/QUANTUM.2020.QTu8B.11} {\emph {\bibinfo {booktitle}
  {OSA Quantum 2.0 Conference}}}\ (\bibinfo  {publisher} {Optica Publishing
  Group},\ \bibinfo {year} {2020})\ p.\ \bibinfo {pages} {QTu8B.11}\BibitemShut
  {NoStop}%
\bibitem [{\citenamefont {Wehner}\ \emph {et~al.}(2018)\citenamefont {Wehner},
  \citenamefont {Elkouss},\ and\ \citenamefont {Hanson}}]{Pirker20182018}%
  \BibitemOpen
  \bibfield  {author} {\bibinfo {author} {\bibfnamefont {S.}~\bibnamefont
  {Wehner}}, \bibinfo {author} {\bibfnamefont {D.}~\bibnamefont {Elkouss}},\
  and\ \bibinfo {author} {\bibfnamefont {R.}~\bibnamefont {Hanson}},\
  }\bibfield  {title} {\bibinfo {title} {Quantum internet: A vision for the
  road ahead},\ }\href {https://doi.org/10.1126/science.aam9288} {\bibfield
  {journal} {\bibinfo  {journal} {Science}\ }\textbf {\bibinfo {volume}
  {362}},\ \bibinfo {pages} {eaam9288} (\bibinfo {year} {2018})}\BibitemShut
  {NoStop}%
\bibitem [{\citenamefont {Branciard}\ \emph {et~al.}(2012)\citenamefont
  {Branciard}, \citenamefont {Cavalcanti}, \citenamefont {Walborn},
  \citenamefont {Scarani},\ and\ \citenamefont {Wiseman}}]{Branciard20122012}%
  \BibitemOpen
  \bibfield  {author} {\bibinfo {author} {\bibfnamefont {C.}~\bibnamefont
  {Branciard}}, \bibinfo {author} {\bibfnamefont {E.~G.}\ \bibnamefont
  {Cavalcanti}}, \bibinfo {author} {\bibfnamefont {S.~P.}\ \bibnamefont
  {Walborn}}, \bibinfo {author} {\bibfnamefont {V.}~\bibnamefont {Scarani}},\
  and\ \bibinfo {author} {\bibfnamefont {H.~M.}\ \bibnamefont {Wiseman}},\
  }\bibfield  {title} {\bibinfo {title} {One-sided device-independent quantum
  key distribution: Security, feasibility, and the connection with steering},\
  }\href {https://doi.org/10.1103/PhysRevA.85.010301} {\bibfield  {journal}
  {\bibinfo  {journal} {Phys. Rev. A}\ }\textbf {\bibinfo {volume} {85}},\
  \bibinfo {pages} {010301} (\bibinfo {year} {2012})}\BibitemShut {NoStop}%
\bibitem [{\citenamefont {Brunner}\ \emph {et~al.}(2014)\citenamefont
  {Brunner}, \citenamefont {Cavalcanti}, \citenamefont {Pironio}, \citenamefont
  {Scarani},\ and\ \citenamefont {Wehner}}]{Brunner14}%
  \BibitemOpen
  \bibfield  {author} {\bibinfo {author} {\bibfnamefont {N.}~\bibnamefont
  {Brunner}}, \bibinfo {author} {\bibfnamefont {D.}~\bibnamefont {Cavalcanti}},
  \bibinfo {author} {\bibfnamefont {S.}~\bibnamefont {Pironio}}, \bibinfo
  {author} {\bibfnamefont {V.}~\bibnamefont {Scarani}},\ and\ \bibinfo {author}
  {\bibfnamefont {S.}~\bibnamefont {Wehner}},\ }\bibfield  {title} {\bibinfo
  {title} {Bell nonlocality},\ }\href
  {https://doi.org/10.1103/RevModPhys.86.419} {\bibfield  {journal} {\bibinfo
  {journal} {Rev. Mod. Phys.}\ }\textbf {\bibinfo {volume} {86}},\ \bibinfo
  {pages} {419} (\bibinfo {year} {2014})}\BibitemShut {NoStop}%
\bibitem [{\citenamefont {Breuer}\ and\ \citenamefont
  {Petruccione}(2002)}]{Breuer&Petruccione02}%
  \BibitemOpen
  \bibfield  {author} {\bibinfo {author} {\bibfnamefont {H.-P.}\ \bibnamefont
  {Breuer}}\ and\ \bibinfo {author} {\bibfnamefont {F.}~\bibnamefont
  {Petruccione}},\ }\href@noop {} {\emph {\bibinfo {title} {The Theory of Open
  Quantum Systems}}}\ (\bibinfo  {publisher} {Oxford Univ. Press},\ \bibinfo
  {year} {2002})\BibitemShut {NoStop}%
\bibitem [{\citenamefont {Vogel}\ and\ \citenamefont {Risken}(1989)}]{Vogel89}%
  \BibitemOpen
  \bibfield  {author} {\bibinfo {author} {\bibfnamefont {K.}~\bibnamefont
  {Vogel}}\ and\ \bibinfo {author} {\bibfnamefont {H.}~\bibnamefont {Risken}},\
  }\bibfield  {title} {\bibinfo {title} {Determination of quasiprobability
  distributions in terms of probability distributions for the rotated
  quadrature phase},\ }\href@noop {} {\bibfield  {journal} {\bibinfo  {journal}
  {Phys. Rev. A.}\ }\textbf {\bibinfo {volume} {40}},\ \bibinfo {pages}
  {2847(R)} (\bibinfo {year} {1989})}\BibitemShut {NoStop}%
\bibitem [{\citenamefont {Bhatia}(1997)}]{Bhatia97}%
  \BibitemOpen
  \bibfield  {author} {\bibinfo {author} {\bibfnamefont {R.}~\bibnamefont
  {Bhatia}},\ }\href@noop {} {\emph {\bibinfo {title} {Matrix Analysis}}}\
  (\bibinfo  {publisher} {Springer International Publishing, Springer},\
  \bibinfo {year} {1997})\BibitemShut {NoStop}%
\bibitem [{\citenamefont {Tomamichel}(2016)}]{Tomamichel16}%
  \BibitemOpen
  \bibfield  {author} {\bibinfo {author} {\bibfnamefont {M.}~\bibnamefont
  {Tomamichel}},\ }\href@noop {} {\emph {\bibinfo {title} {Quantum Information
  Processing with Finite Resources}}}\ (\bibinfo  {publisher} {Springer
  International Publishing, Springer},\ \bibinfo {year} {2016})\BibitemShut
  {NoStop}%
\end{thebibliography}%

\end{document}